\documentclass[aps,prb,twocolumn]{revtex4-1}

\usepackage[english]{babel}
\usepackage[utf8]{inputenc}
\usepackage{amsmath}
\usepackage{amssymb}
\usepackage[caption=false]{subfig}
\usepackage{amssymb}
\usepackage{epsfig}
\usepackage{graphicx}
\usepackage{amsmath}
\usepackage{array,color}
\usepackage{natbib}

\usepackage{soul}
\usepackage{textcomp}
\usepackage[T1]{fontenc}

\usepackage[usenames,dvipsnames]{xcolor}
\definecolor{forestgreen}{rgb}{0.11,0.54,0.15}
\definecolor{purple}{rgb}{0.62,0.10,0.96}
\definecolor{dockerblue}{rgb}{0.11,0.56,0.98}
\definecolor{freeblue}{rgb}{0.25,0.41,0.88}

\usepackage[pdftex,plainpages=false,colorlinks=true,linkcolor=Red, citecolor=blue, urlcolor=blue]{hyperref}

\usepackage{latexsym}
\usepackage{url}
\usepackage{xcolor}

\usepackage{bm}

\usepackage{etoolbox}

\usepackage{appendix}

\begin{document}

\title{Matrix moments of the diffusion tensor distribution}

\author{A. Reymbaut$^{1,2}$}
\email{alexis.reymbaut@fkem1.lu.se}

\affiliation{
$^1$Department of Physical Chemistry, Lund University, Lund, Sweden\\
$^2$Random Walk Imaging AB, Lund, Sweden
}

\date{\today}

\begin{abstract}
\textbf{Purpose:} To facilitate the implementation/validation of signal representations and models using parametric matrix-variate distributions to approximate the diffusion tensor distribution (DTD) $\mathcal{P}(\mathbf{D})$. \\
\textbf{Theory:} We establish practical mathematical tools, the matrix moments of the DTD, enabling to compute the mean diffusion tensor and covariance tensor associated with any parametric matrix-variate DTD whose moment-generating function is known. As a proof of concept, we apply these tools to the non-central matrix-variate Gamma (nc-mv-Gamma) distribution, whose covariance tensor was so far unknown, and design a new signal representation capturing intra-voxel heterogeneity \textit{via} a single nc-mv-Gamma distribution: the matrix-variate Gamma approximation. \\
\textbf{Methods:} Furthering this proof of concept, we evaluate the matrix-variate Gamma approximation \textit{in silico} and \textit{in vivo}, in a human-brain `tensor-valued' diffusion MRI dataset. \\
\textbf{Results:} The matrix-variate Gamma approximation fails to capture the heterogeneity arising from orientation dispersion and from simultaneous variances in the trace (size) and anisotropy (shape) of the underlying diffusion tensors, which is explained by the structure of the covariance tensor associated with the nc-mv-Gamma distribution. \\
\textbf{Conclusion:} The matrix moments promote a more widespread use of matrix-variate distributions as plausible approximations of the DTD by alleviating their intractability, thereby facilitating the design/validation of matrix-variate microstructural techniques. 
\end{abstract}

\maketitle

\footnotetext[1]{\textbf{Abbreviations used:} MRI, magnetic resonance imaging; dMRI, diffusion MRI; MADCO, marginal distributions constrained optimization; DIAMOND, distribution of anisotropic microstructural environments in diffusion-compartment imaging; mv-Gaussian, matrix-variate Gaussian; nc-mv-Gamma, non-central matrix-variate Gamma; SNR, signal-to-noise ratio; CSF, cerebrospinal fluid; GM, grey matter; WM, white matter; mv-Gamma, matrix-variate Gamma approximation; Cov, covariance tensor approximation; DEC, directionally encoded color.}


\newpage

\section{Introduction}
\label{Sec_Intro}

Diffusion MRI (dMRI), which captures the translational motion of water molecules diffusing in biological tissue,~\citep{LeBihan:1990, LeBihan:1992, Basser:1994, Mattiello:1994, Mattiello:1997, Jones:2010} has provided critical sensitivity to tissue microstructure \textit{in vivo}. Nevertheless, microstructural dMRI studies have been impeded by the lack of specificity of the measured diffusion signal, which is only sensitive to the voxel-averaged diffusion profile. Given that typical cubic-millimeter dMRI voxels comprise multiple cell types and the extra-cellular space,~\citep{Stanisz:1997,Norris:2001,Sehy:2002,Minati:2007,Mulkern:2009} the dMRI signal probes a collection of microscopic diffusion profiles over a specific observational time-scale that depends on the choice of experimental time parameters. A common description~\citep{Jian:2007} of the interplay between the voxel content and the measured dMRI signal $\mathcal{S}$, dubbed the "diffusion tensor distribution" (DTD) description, is attained by considering a `snapshot' of the combined non-Gaussian diffusion effects of restriction~\citep{Woessner:1963} and exchange~\citep{Johnson:1993,Li_Springer:2019} at this given observational time-scale, and by approximating the signal decay as a continuous weighted sum of exponential decays,~\citep{Alexander:2001,Tuch:2002,Yablonskiy:2003,Kroenke:2004,Jespersen:2007,Leow:2009,Pasternak:2009,Wang:2011,Fieremans:2011,Zhang_NODDI:2012,Jelescu:2016,Kaden:2016,Westin:2016,Scherrer_DIAMOND:2016,Scherrer_aDIAMOND:2017,Lampinen_CODIVIDE:2017,Reisert:2017,Novikov_on_modeling:2018,Novikov_WMSM:2018,Rensonnet:2018} yielding
\begin{equation}
\frac{\mathcal{S}(\mathbf{b})}{\mathcal{S}_0} = \int_{\mathrm{Sym}^+(3)} \mathcal{P}(\mathbf{D}) \exp(-\mathbf{b}:\mathbf{D})\, \mathrm{d}\mathbf{D} =\left\langle\exp(-\mathbf{b}:\mathbf{D})\right\rangle \, ,
\label{Eq_signal_tensor_distribution}
\end{equation}
where $\mathbf{b}$ is the symmetric diffusion-encoding b-tensor from tensor-valued diffusion encoding,~\citep{Eriksson:2013,Westin:2014,Eriksson:2015,Westin:2016,Topgaard:2017,Topgaard_dim_rand_walks:2019} $\mathcal{S}_0=\mathcal{S}(\mathbf{b}=\mathbf{0})$ is the non diffusion-weighted signal, and $\mathcal{P}(\mathbf{D})$ is the intra-voxel distribution of apparent diffusion tensors $\mathbf{D}$. While $\mathrm{Sym}^{+}(3)$ denotes the space of real symmetric positive definite $3\times 3$ tensors, ":" is the Frobenius inner product, and $\langle\,\cdot\,\rangle$ corresponds to the voxel-scale average, \textit{i.e.} the average computed over the voxel content. 
The validity of the DTD description is discussed in Appendix~\ref{Sec_validity_DTD}.

Even within the convenient description formulated in Equation~\ref{Eq_signal_tensor_distribution}, it remains a challenge to either estimate the distribution $\mathcal{P}(\mathbf{D})$ as a whole or to estimate its main features, also called "statistical descriptors".~\citep{Reymbaut_accuracy_precision:2020} While non-parametric techniques have been developed to retrieve the entire DTD from dMRI data, such as the marginal distributions constrained optimization (MADCO),~\citep{Benjamini:2016,Benjamini:2018,Benjamini:2020} multidimensional correlation spectroscopic imaging,~\citep{Kim:2017,Kim:2020} and Monte-Carlo signal inversions,~\citep{deAlmeidaMartins_Topgaard:2016, deAlmeidaMartins_Topgaard:2018, Topgaard:2019, deAlmeidaMartins:2020} an alternative approach consists in approximating $\mathcal{P}(\mathbf{D})$ with a plausible parametric functional form whose parameters can be fitted against the acquired signal. This parametric approach encompasses diffusion tensor imaging (DTI),~\citep{Basser:1994} signal representations based on normal,~\citep{Yablonskiy:2003, Jensen:2005, Kiselev:2012, Renaud:2015, Mohanty:2018} log-normal~\citep{Hakansson:2000,Williamson:2016} and Gamma~\citep{Roding:2012, Lasic:2014, Williamson:2016} distributions of diffusivities, models \citep{Assaf:2004, Assaf_CHARMED:2005, Assaf:2008, Jbabdi:2012, Zhang_NODDI:2012, Lampinen_CODIVIDE:2017, Novikov_WMSM:2018} and higher-than-second-order truncated cumulant expansions.~\citep{Ning:2018, Ning:2020} In particular, two-term cumulant expansions of the low b-value diffusion signal are equivalent to considering a normal distribution of diffusivities, as detailed in previous work.~\citep{Yablonskiy:2003} The parametric approach has been extended to distributions of diffusion tensors in three ways:
\begin{itemize}
\item as mixtures of Wishart distributions~\citep{Muirhead_Book:1982, Gupta_Nagar_Book:2000, Anderson_Book:2003} used to capture the intra-voxel orientation distribution function in a spherical deconvolution approach,~\citep{Tournier:2004,Tournier:2007} as found in Ref.~\onlinecite{Jian:2007}.
\item as voxel-scale matrix-variate Gaussian (mv-Gaussian) distributions~\citep{Muirhead_Book:1982, Gupta_Nagar_Book:2000, Anderson_Book:2003} within the original work of Refs.~\onlinecite{Basser_Pajevic:2003, Pajevic_Basser:2003} and the covariance tensor approximation of Ref.~\onlinecite{Westin:2016}, describing two-term cumulant expansions of the DTD.
\item as compartmental non-central matrix-variate Gamma (nc-mv-Gamma) distributions~\citep{Gupta_Nagar_Book:2000} within the distribution of anisotropic microstructural environments in diffusion-compartment imaging (DIAMOND) model,~\citep{Scherrer_DIAMOND:2016,Scherrer_aDIAMOND:2017, Reymbaut_arxiv_Magic_DIAMOND:2020} wherein each sub-voxel anisotropic compartment is described by a nc-mv-Gamma distribution.
\end{itemize}
The adjective "matrix-variate" refers to functions of matrix argument. On the one hand, while spherical deconvolution~\citep{Tournier:2004,Tournier:2007} relies on a pre-specified convolution kernel that may disagree with the underlying microstructure,~\citep{Parker:2013,Tax:2014} the mv-Gaussian distribution allows for unphysical negative definite diffusion tensors and has been shown to exhibit biases when estimating the DTD's statistical descriptors for certain tissue configurations.~\citep{Reymbaut_accuracy_precision:2020} It is worth mentioning that recent work has aimed to limit this distribution to the space of positive semidefinite tensors.~\citep{Magdoom:2020} On the other hand, the nc-mv-Gamma distribution is by definition restricted to the space of positive definite tensors. However, it appears to be intractable for defining statistical descriptors straightforwardly comparable to those obtained by other non-parametric or parametric techniques, unlike the mv-Gaussian distribution.~\citep{Westin:2016, Magdoom:2020} This lack of tractability, inherent to most matrix-variate distributions (except for the mv-Gaussian distribution), hinders the cross-validation of current signal representations/models relying on such mathematical objects,~\citep{Scherrer_DIAMOND:2016,Scherrer_aDIAMOND:2017, Reymbaut_arxiv_Magic_DIAMOND:2020} and impedes the design of novel matrix-variate parametric techniques.

In this work, we derive general tools facilitating the implementation and validation of any parametric matrix-variate functional choice for $\mathcal{P}(\mathbf{D})$: the matrix moments of the diffusion tensor distribution. These matrix moments enable the computation of the mean diffusion tensor and covariance tensor associated with a given parametric approximation of $\mathcal{P}(\mathbf{D})$, from which common statistical descriptors of the DTD can be estimated. In turn, these descriptors can be used to assess the limitations of the parametric approximation, and to compare it with other techniques on the basis of estimating identical sets of statistical descriptors. As a proof of concept, we apply these matrix moments to the non-central matrix-variate Gamma distribution, thereby developing a new signal representation wherein the voxel content is described by a single nc-mv-Gamma DTD: the "matrix-variate Gamma approximation". In addition, the definitions of the statistical descriptors within this approximation can be used in the aforementioned DIAMOND model to quantify fiber-specific diffusion features in a way that is comparable to other fiber-specific methods.~\citep{Assaf:2004, Assaf_CHARMED:2005, Reymbaut_arxiv_MC_DPC:2020} Finally, we evaluate this approximation \textit{in vivo} and \textit{in silico}. Note that while the matrix moments of matrix-variate distributions were already introduced and computed for the mv-Gaussian and Wishart distributions in Ref.~\onlinecite{Kollo_von_Rosen_book:2006}, the present work applies them for the first time to the dMRI field and the nc-mv-Gamma distribution.

In Section~\ref{Sec_theory}, we first detail the mathematical steps leading to the formulation of the matrix moments in Section~\ref{Sec_matrix_moments_D_C}, and then apply these tools to the nc-mv-Gamma distribution in Section~\ref{Sec_mv_Gamma} before establishing the matrix-variate Gamma approximation in Section~\ref{Sec_mv_Gamma_approximation}. In Section~\ref{Sec_Methods}, we review the methods used for the \textit{in vivo} and \textit{in silico} evaluations of this approximation, and for its \textit{in silico} comparison with the covariance tensor approximation of Ref.~\onlinecite{Westin:2016}. We present our results in Section~\ref{Sec_Results}, discuss them in Section~\ref{Sec_Discussion}, and conclude in Section~\ref{Sec_Conclusions}. 



\section{Theory}
\label{Sec_theory}

\subsection{Estimating statistical descriptors from the mean diffusion tensor and the covariance tensor}
\label{Sec_diffusion_metrics}

Parametrizing a given axisymmetric diffusion tensor $\mathbf{D}$ by its axial diffusivity $D_\parallel$, radial diffusivity $D_\perp$ and orientation $(\theta, \phi)$, one defines the isotropic diffusivity $D_\mathrm{iso} = (D_\parallel + 2D_\perp)/3$, normalized anisotropy $D_\Delta = (D_\parallel - D_\perp)/(3D_\mathrm{iso})\in[-0.5,1]$ and anisotropic diffusivity $D_\mathrm{aniso} = D_\mathrm{iso}D_\Delta = (D_\parallel - D_\perp)/3$.~\citep{Haeberlen:1976} Within this parametrization, common statistical descriptors of the DTD are given by the mean diffusivity $\mathrm{E}[D_\mathrm{iso}]$, the variance of isotropic diffusivities $\mathrm{V}[D_\mathrm{iso}]$ and the mean squared anisotropic diffusivity $\mathrm{E}[D_\mathrm{aniso}^2]$. The mean squared anisotropic diffusivity can also be normalized as the normalized mean squared anisotropy $\tilde{\mathrm{E}}[D_\mathrm{aniso}^2] = \mathrm{E}[D_\mathrm{aniso}^2]/ \mathrm{E}[D_\mathrm{aniso}]^2$. We retained the notations $\mathrm{E}[\, \cdot\, ]$ for the voxel-scale expectation and $\mathrm{V}[\, \cdot\, ]$ for the voxel-scale variance to be consistent with previous works.~\citep{deAlmeidaMartins_Topgaard:2018, Topgaard:2019, Reymbaut_accuracy_precision:2020, deAlmeidaMartins:2020, Reymbaut_arxiv_MC_DPC:2020} These descriptors can be related to other measures derived in the dMRI-microstructure literature. $\mathrm{E}[\mathit{D}_\mathrm{iso}]$ is identical to the mean diffusivity (MD). $\tilde{\mathrm{E}}[\mathit{D}_\mathrm{aniso}^2]$ carries similar information to the microscopic anisotropy index (MA),\cite{Lawrenz:2010} the fractional eccentricity (FE),\cite{Jespersen:2013} the microscopic anisotropy ({\textmu}A),\cite{Shemesh:2016,Ianus:2018} the normalized difference between second moments $\Delta\tilde{\mu}_2$ and microscopic fractional anisotropy ({\textmu}FA),\cite{Lasic:2014,Shemesh:2016} the anisotropic variance $\mathit{V}_\mathrm{A}$ and anisotropic mean kurtosis (MK$_\mathrm{A}$),\cite{Szczepankiewicz:2015,Szczepankiewicz:2016} and the microscopic anisotropy $\mathit{C}_\mu$.\cite{Westin:2016} $\mathrm{V}[\mathit{D}_\mathrm{iso}]$ yields similar information to the isotropic second moment $\mu_2^\mathrm{iso}$,\cite{Lasic:2014} the isotropic variance $\mathit{V}_\mathrm{I}$ and isotropic mean kurtosis (MK$_\mathrm{I}$),\cite{Szczepankiewicz:2015,Szczepankiewicz:2016} and the normalized isotropic variance $\mathit{C}_\mathrm{MD}$.\cite{Westin:2016}


The work of Ref.~\onlinecite{Westin:2016} establishes how to compute statistical descriptors of the DTD $\mathcal{P}(\mathbf{D})$ from its mean diffusion tensor $\langle \mathbf{D}\rangle$ and covariance tensor $\mathbb{C}=\langle\mathbf{D}^{\otimes 2}\rangle-\langle\mathbf{D}\rangle^{\otimes\!\,2}$. "$\otimes$" denotes the outer tensor product, with the short-hand notation $\mathbf{D}^{\otimes 2} = \mathbf{D}\otimes\mathbf{D}$. Let us introduce the Mandel notation in which a $3\times 3$ symmetric tensor $\bm{\Lambda}$ writes as an equivalent $6\times 1$ column vector following~\citep{Mandel:1965}
\begin{align}
\bm{\Lambda} & = 
\begin{pmatrix}
\lambda_{11} & \lambda_{12} & \lambda_{13}\\
\cdot & \lambda_{22} & \lambda_{23} \\ 
\cdot & \cdot & \lambda_{33} 
\end{pmatrix} \nonumber \\
& \equiv 
\begin{pmatrix}
\lambda_{11} & \lambda_{22} & \lambda_{33} & \sqrt{2}\,\lambda_{23} & \sqrt{2}\,\lambda_{13}& \sqrt{2}\,\lambda_{12}
\end{pmatrix}^\text{T} \nonumber \\
& = \bm{\Lambda}_\text{Mandel} \,,
\end{align}
where "$\mathrm{T}$" indicates vector/matrix transposition. In Mandel notation, the $9\times 9$ outer tensor product $\bm{\Lambda}_1 \otimes \bm{\Lambda}_2$ of two $3\times 3$ symmetric tensors is equivalent to a $6\times 6$ tensor according to
\begin{equation}
\bm{\Lambda}_1 \otimes \bm{\Lambda}_2 \equiv \bm{\Lambda}_{1,\text{Mandel}} \cdot \bm{\Lambda}_{2,\text{Mandel}}^\text{T}\,,
\label{Eq_Mandel}
\end{equation}
where "$\cdot$" is the standard vector/matrix multiplication. We now omit the equivalent sign "$\equiv$". Following Ref.~\onlinecite{Westin:2016}, one defines $\mathbf{E}_{\mathrm{iso}} =\mathbf{I}_3/3$ and $\mathbb{E}_{\mathrm{iso}} = \mathbf{I}_6/3$ (with the $n\times n$ identity matrix $\mathbf{I}_n$), and builds the bulk and shear modulus tensors
\begin{align}
\mathbb{E}_\mathrm{bulk} = \mathbf{E}_{\mathrm{iso}}^{\otimes 2} & = \frac{1}{9}
\begin{pmatrix}
1 & 1 & 1 & 0 & 0 & 0 \\
1 & 1 & 1 & 0 & 0 & 0 \\
1 & 1 & 1 & 0 & 0 & 0 \\
0 & 0 & 0 & 0 & 0 & 0 \\ 
0 & 0 & 0 & 0 & 0 & 0 \\
0 & 0 & 0 & 0 & 0 & 0
\end{pmatrix} \;, \\
\mathbb{E}_\mathrm{shear} = \mathbb{E}_{\mathrm{iso}}-\mathbb{E}_\mathrm{bulk} & = \frac{1}{9}
\begin{pmatrix}
2 & -1 & -1 & 0 & 0 & 0 \\
-1 & 2 & -1 & 0 & 0 & 0 \\
-1 & -1 & 2 & 0 & 0 & 0 \\
0 & 0 & 0 & 3 & 0 & 0 \\
0 & 0 & 0 & 0 & 3 & 0 \\ 
0 & 0 & 0 & 0 & 0 & 3
\end{pmatrix} 
\end{align}
by analogy with the stress tensor in mechanics. From these tensors, one obtains the aforementioned statistical descriptors as 
\begin{align}
\mathrm{E}[\mathit{D}_{\mathrm{iso}}] & = \langle \mathbf{D}\rangle : \mathbf{E}_{\mathrm{iso}},  \nonumber \\
\mathrm{V}[\mathit{D}_{\mathrm{iso}}] & = \mathbb{C}:\mathbb{E}_\mathrm{bulk}, \label{Eq_statistical_descriptors_D_C} \\
\mathrm{E}[\mathit{D}_\mathrm{aniso}^2] & = \frac{(\mathbb{C}+\langle\mathbf{D}\rangle^{\otimes 2}):\mathbb{E}_\mathrm{shear}}{2} = \frac{\langle\mathbf{D}^{\otimes 2}\rangle:\mathbb{E}_\mathrm{shear}}{2}\,, \nonumber
\end{align}
among other statistical descriptors that also depend on $\langle \mathbf{D}\rangle $ and $\mathbb{C}$.~\cite{Westin:2016,Magdoom:2020} 


\subsection{Matrix moments of the diffusion tensor distribution}
\label{Sec_matrix_moments}

\subsubsection{Moment-generating function}

Moment-generating functions of scalar distributions are commonly used to access the moments of these distributions. In the case of a scalar distribution of diffusivities $\mathcal{P}(D)$, its moment-generating function reads
\begin{equation}
\mathcal{M}(z) = \int_0^{+\infty} \mathcal{P}(D)\, \exp(zD)\, \mathrm{d}D = \langle \exp(zD)\rangle\, ,
\label{Eq_moment_generating_function_scalar}
\end{equation}
with $z\in\mathbb{R}$ so that $\mathcal{M}(z)$ converges, and the average $\langle \,\cdot\,\rangle$ taken over $\mathcal{P}(D)$. While the characteristic function $\varphi(z)=\mathcal{M}(iz)$ is properly defined for all $z\in\mathbb{R}$ (because it corresponds to the integral of a bounded function on a space of finite measure), the convergence of Equation~\ref{Eq_moment_generating_function_scalar} may be limited to certain ranges of values for $z$. The link between the moment-generating function $\mathcal{M}(z)$ and the raw moments $m_n$ of $\mathcal{P}(D)$, with $n\in\mathbb{N}$, is manifest upon Taylor-expanding the exponential in Equation~\ref{Eq_moment_generating_function_scalar} and computing the derivatives of $\mathcal{M}(z)$:
\begin{equation}
m_n = \langle D^n \rangle = \left. \frac{\mathrm{d}^n\mathcal{M}(z)}{\mathrm{d}z^n} \right\vert_{z=0}\, .
\label{Eq_scalar_moments}
\end{equation}
In particular, $m_0 = 1$, $m_1 = \langle D\rangle$ and $m_2 = \langle D^2\rangle$, so that the variance of diffusivities in $\mathcal{P}(D)$ is given by $m_2 - m_1^2 = \langle D^2\rangle - \langle D\rangle^2$.

For a diffusion tensor distribution $\mathcal{P}(\mathbf{D})$, its matrix-variate moment-generating function $\mathcal{M}(\mathbf{Z})$ is defined as~\citep{Gupta_Nagar_Book:2000}
\begin{equation}
\mathcal{M}(\mathbf{Z}) = \int_{\mathrm{Sym}^+(3)} \mathcal{P}(\mathbf{D}) \exp(\mathbf{Z}:\mathbf{D})\, \mathrm{d}\mathbf{D} = \left\langle\exp(\mathbf{Z}:\mathbf{D})\right\rangle
\label{Eq_moment_generating_function}
\end{equation}
for $\mathbf{Z}\in\mathrm{Sym}(3)$ such that $\mathcal{M}(\mathbf{Z})$ converges, $\mathrm{Sym}(3)$ being the space of real symmetric $3\times 3$ tensors. Note that in the mathematics literature, the Frobenius inner product ":" is usually denoted by its trace definition, \textit{i.e.} $\mathbf{Z}:\mathbf{D} = \mathrm{Tr}(\mathbf{Z}^{\mathrm{T}}\cdot\mathbf{D})$, equivalent to $\mathrm{Tr}(\mathbf{Z}\cdot\mathbf{D})$ by symmetry of $\mathbf{Z}$. It is clear from Equations~\ref{Eq_signal_tensor_distribution} and \ref{Eq_moment_generating_function} that the dMRI signal associated with any parametric approximation of $\mathcal{P}(\mathbf{D})$ can be obtained \textit{via} its moment-generating function $\mathcal{M}(z)$ (available in various matrix-variate statistics books~\citep{Muirhead_Book:1982, Gupta_Nagar_Book:2000, Anderson_Book:2003}) as~\citep{Reymbaut_arxiv_Magic_DIAMOND:2020}
\begin{equation}
\frac{\mathcal{S}(\mathbf{b})}{\mathcal{S}_0} = \mathcal{M}(-\mathbf{b})\, .
\label{Eq_moment_generating_function_link_signal}
\end{equation}
By analogy with the link between the derivatives of the moment-generating function Equation~\ref{Eq_moment_generating_function_scalar} and the scalar moments Equation~\ref{Eq_scalar_moments} of any scalar distribution, we show in Section~\ref{Sec_matrix_moments_D_C} that the first-order and second-order matrix derivatives of $\mathcal{M}(\mathbf{Z})$ are linked to the matrix moments $\langle\mathbf{D}\rangle$ and $\langle\mathbf{D}^{\otimes 2}\rangle$.~\citep{Kollo_von_Rosen_book:2006}

\subsubsection{Matrix calculus and layout convention}
\label{Sec_matrix_calculus}

Let us introduce a $p\times q$ matrix $\mathbf{X}$ of matrix elements $(x_{ij})_{1\leq i \leq p, 1\leq j \leq q}$. In numerator layout convention,~\citep{Wiki_tensor_calculus} the first-order matrix derivative of a scalar-valued matrix-variate function $g$ of $\mathbf{X}$ with respect to $\mathbf{X}$ is given by~\citep{Turnbull:1928, Turnbull:1930, Turnbull:1931, Dwyer_MacPhail:1948, Selby_Weast_book:1967, MacRae:1974, Magnus_Neudecker:1999, Kollo_von_Rosen_book:2006,Magnus:2010}
\begin{equation}
\frac{\partial g}{\partial \mathbf{X}} = 
\begin{pmatrix}
\frac{\partial g}{\partial x_{11}} & \frac{\partial g}{\partial x_{12}} & \cdots & \frac{\partial g}{\partial x_{1q}} \\
\frac{\partial g}{\partial x_{21}} & \frac{\partial g}{\partial x_{22}} & \cdots & \frac{\partial g}{\partial x_{2q}} \\
\vdots & \vdots & \ddots & \vdots \\
\frac{\partial g}{\partial x_{p1}} & \frac{\partial g}{\partial x_{p2}} & \cdots & \frac{\partial g}{\partial x_{pq}}
\end{pmatrix} \, .
\label{Eq_first_derivative_matrix_calculus}
\end{equation}
In order to compute first-order matrix derivatives, one can refer to the resources found in Refs.~\onlinecite{Laue:2018,Laue_website}, primarily designed for neural networks in the machine-learning field. However, these references use a mixed layout convention that is equivalent to taking the transpose of Equation~\ref{Eq_first_derivative_matrix_calculus} as result for $\partial g/\partial \mathbf{X}$ (denominator layout convention).~\citep{Wiki_tensor_calculus} Given that this present work will use the numerator layout convention, the results yielded by Refs.~\onlinecite{Laue:2018,Laue_website} for $\partial g/\partial \mathbf{X}$ have to be transposed to match our problem at hand.

As for the second-order matrix derivative of $g$ with respect to $\mathbf{X}$, $\partial^2 g/\partial \mathbf{X}^2$, it yields a fourth-order tensor for which no general notation convention is widely agreed upon.~\citep{Wiki_tensor_calculus,Kollo_von_Rosen_book:2006} Nonetheless, matrix-calculus rules do exist to compute it.~\citep{Brewer:1978,Kollo_von_Rosen_book:2006} In particular, one can use the scalar product rule~\citep{Brewer:1978}
\begin{equation}
\frac{\partial[f(\mathbf{X})\times\mathbf{F}(\mathbf{X})]}{\partial\mathbf{X}}=\frac{\partial\!\,f}{\partial\mathbf{X}}\otimes\mathbf{F}(\mathbf{X})+f(\mathbf{X})\times\frac{\partial\mathbf{F}}{\partial\mathbf{X}} \, ,
\label{Eq_scalar_product_rule}
\end{equation}
where $f$ is a scalar-valued matrix-variate function of $\mathbf{X}$ and $\mathbf{F}$ is a matrix-valued matrix-variate function of $\mathbf{X}$, to compute the second-order matrix derivative of $g$ with respect to $\mathbf{X}$ given that its first-order derivative can be rewritten as $\partial g/\partial \mathbf{X} = f(\mathbf{X})\times\mathbf{F}(\mathbf{X})$. "$\times$" denotes the scalar multiplication, used to render certain equations unambiguous. 

\begin{widetext}

\subsubsection{Matrix moments, mean diffusion tensor and covariance tensor}
\label{Sec_matrix_moments_D_C}

Using the aforementioned relationship $\mathbf{Z}:\mathbf{D} = \mathrm{Tr}(\mathbf{Z}\cdot\mathbf{D})$ and common matrix-calculus rules,~\citep{Brewer:1978,Kollo_von_Rosen_book:2006} the first-order matrix derivative of the moment-generating function $\mathcal{M}(\mathbf{Z})$ in Equation~\ref{Eq_moment_generating_function} writes
\begin{equation}
\frac{\partial\mathcal{M}}{\partial\mathbf{Z}} =\frac{\partial\left\langle\exp[\mathrm{Tr}(\mathbf{Z}\cdot\mathbf{D})]\right\rangle}{\partial\mathbf{Z}} = \left\langle\frac{\partial \exp[\mathrm{Tr}(\mathbf{Z}\cdot\mathbf{D})]}{\partial \mathbf{Z}}\right\rangle = \left\langle\exp[\mathrm{Tr}(\mathbf{Z}\cdot\mathbf{D})]\times\mathbf{D}\right\rangle \, ,
\label{Eq_interm_10}
\end{equation}
so that the mean diffusion tensor of $\mathcal{P}(\mathbf{D})$ is given by
\begin{equation}
\langle\mathbf{D}\rangle=\left.\frac{\partial\mathcal{M}}{\partial\mathbf{Z}}\right|_{\mathbf{Z}=\mathbf{0}} \, .
\label{Eq_first_moment}
\end{equation}
As for the second-order matrix derivative of $\mathcal{M}(\mathbf{Z})$, one can combine the scalar product rule of Equation~\ref{Eq_scalar_product_rule} and Equation~\ref{Eq_interm_10}, to obtain
\begin{equation}
\frac{\partial^2\mathcal{M}}{\partial\mathbf{Z}^2} = \frac{\partial\;}{\partial\mathbf{Z}}\, \frac{\partial\mathcal{M}}{\partial\mathbf{Z}} = \left\langle\frac{\partial [\exp[\mathrm{Tr}(\mathbf{Z}\cdot\mathbf{D})]\times\mathbf{D}]}{\partial\mathbf{Z}}\right\rangle = \bigg\langle\underbrace{\left[\frac{\partial\exp[\mathrm{Tr}(\mathbf{Z}\cdot\mathbf{D})]}{\partial\mathbf{Z}}\right]}_{\exp\left[\mathrm{Tr}(\mathbf{Z}\cdot\mathbf{D})\right]\times\mathbf{D}}\otimes\,\mathbf{D}+\exp[\mathrm{Tr}(\mathbf{Z}\cdot\mathbf{D})]\times\underbrace{\frac{\partial\mathbf{D}}{\partial \mathbf{Z}}}_{\mathbf{0}}\bigg\rangle\, ,
\end{equation}
giving
\begin{equation}
\frac{\partial^2\mathcal{M}}{\partial\mathbf{Z}^2} = \left\langle\exp\!\left[\mathrm{Tr}(\mathbf{Z}\cdot\mathbf{D})\right]\times \mathbf{D}\otimes\mathbf{D}\right\rangle \, ,
\end{equation}
so that 
\begin{equation}
\langle\mathbf{D}^{\otimes 2}\rangle = \left.\frac{\partial^2\mathcal{M}}{\partial\mathbf{Z}^2}\right|_{\mathbf{Z}=\mathbf{0}}\, .
\label{Eq_second_derivative}
\end{equation}
Consequently, the covariance tensor of $\mathcal{P}(\mathbf{D})$ writes
\begin{equation}
\mathbb{C}=\langle\mathbf{D}^{\otimes 2}\rangle-\langle\mathbf{D}\rangle^{\otimes\!\,2}=\left.\frac{\partial^2\mathcal{M}}{\partial\mathbf{Z}^2}\right|_{\mathbf{Z}=\mathbf{0}}-\left.\frac{\partial\mathcal{M}}{\partial\mathbf{Z}}\right|_{\mathbf{Z}=\mathbf{0}}^{\otimes\!\,2} \, .
\label{Eq_second_moment}
\end{equation}
Notice the resemblance between Equations~\ref{Eq_first_moment} and \ref{Eq_second_derivative}, and Equation~\ref{Eq_scalar_moments}. Similar expressions of Equations~\ref{Eq_first_moment} and \ref{Eq_second_derivative} can be found in Ref.~\onlinecite{Kollo_von_Rosen_book:2006}, computed for the mv-Gaussian and Wishart distributions in particular, and are here applied for the first time in the dMRI field. The formulations Equations~\ref{Eq_first_moment} and \ref{Eq_second_moment} for $\langle \mathbf{D}\rangle$ and $\mathbb{C}$ are validated in Appendix~\ref{Sec_validation} on the mv-Gaussian distribution featured in Refs.~\onlinecite{Basser_Pajevic:2003, Pajevic_Basser:2003,Westin:2016}. We emphasize that these equations enable the direct computation of the statistical descriptors in Section~\ref{Sec_diffusion_metrics} according to Equation~\ref{Eq_statistical_descriptors_D_C} for any parametric approximation of $\mathcal{P}(\mathbf{D})$ whose moment-generating function is known.

\subsection{Application to the non-central matrix-variate Gamma distribution}

\subsubsection{Matrix moments of the non-central matrix-variate Gamma distribution}
\label{Sec_mv_Gamma}

The idea of describing the diffusion profile of anisotropic diffusion compartments \textit{via} a matrix-variate Gamma distribution~\citep{Gupta_Nagar_Book:2000} is originally found in the DIAMOND model,~\citep{Scherrer_DIAMOND:2016,Scherrer_aDIAMOND:2017} which considers a free-water compartment accompanied by up to three anisotropic compartments, each described separately by a matrix-variate Gamma distribution. While Ref.~\onlinecite{Scherrer_DIAMOND:2016} uses a central matrix-variate Gamma distribution, Ref.~\onlinecite{Scherrer_aDIAMOND:2017} extends this statistical description by using a non-central matrix-variate Gamma (nc-mv-Gamma) distribution
\begin{equation}
\mathcal{P}_\Gamma(\mathbf{D})=\frac{\mathrm{Det}(\mathbf{D})^{\kappa-2}}{\mathrm{Det}(\bm{\Psi})^\kappa\,\Gamma_3(\kappa)}\,\exp\!\left[-\mathrm{Tr}(\bm{\Theta}+\bm{\Psi}^{-1}\cdot\mathbf{D})\right]\,\mathcal{F}_{0,1}(\kappa,\bm{\Theta}\cdot\bm{\Psi}^{-1}\cdot\mathbf{D})\, ,
\label{Eq_nc_mv_Gamma_dist}
\end{equation}
where $\kappa\in\;]1,+\infty[$ is the shape parameter, $\bm{\Psi}\in\mathrm{Sym}^+(3)$ is the scale parameter, and $\bm{\Theta}\in \mathrm{Sym}(3)$ is the noncentrality parameter (setting $\bm{\Theta} = \mathbf{0}$ gives the central matrix-variate Gamma distribution). $\Gamma_3(\kappa)=\pi^{3/2}\prod_{j=1}^3\Gamma(\kappa-(j-1)/2)$ is the multivariate Gamma function and $\mathcal{F}_{0,1}$ is the hypergeometric (Bessel) function of matrix argument of order $(0,1)$. 

Little is known about the nc-mv-Gamma distribution (compared to the mv-Gaussian and Wishart distributions),~\citep{Muirhead_Book:1982, Gupta_Nagar_Book:2000, Anderson_Book:2003, Kollo_von_Rosen_book:2006} except for the expression of its mean tensor,~\citep{Scherrer_aDIAMOND:2017} $\langle \mathbf{D}\rangle=\bm{\Psi}\cdot[\kappa \mathbf{I}_3+\bm{\Theta}]$, that we retrieve in Equation~\ref{Eq_average_diffusion_tensor_Gamma}, and for its moment-generating function $\mathcal{M}_\Gamma(\mathbf{Z})$, defined for $\mathbf{Z}\in\mathrm{Sym}(3)$ such that $(\mathbf{I}_3-\mathbf{Z}\cdot\bm{\Psi})\in\mathrm{Sym}^+(3)$ as~\citep{Gupta_Nagar_Book:2000}
\begin{equation}
\mathcal{M}_\Gamma(\mathbf{Z}) = \left[\mathrm{Det}(\mathbf{I}_3-\mathbf{Z}\cdot\bm{\Psi})\right]^{-\kappa}\exp\!\left[\mathrm{Tr}\!\left(\left[(\mathbf{I}_3-\mathbf{Z}\cdot\bm{\Psi})^{-1}-\mathbf{I}_3\right]\cdot\bm{\Theta} \right)\right] \, .
\label{Eq_moment_generating_function_gamma}
\end{equation}
As shown in Ref.~\onlinecite{Reymbaut_arxiv_Magic_DIAMOND:2020}, and according to Equation~\ref{Eq_moment_generating_function_link_signal}, the diffusion signal decay associated with the nc-mv-Gamma distribution expresses as
\begin{equation}
\frac{\mathcal{S}_\Gamma (\mathbf{b})}{\mathcal{S}_0}=\mathcal{M}_\Gamma(-\mathbf{b})=[\mathrm{Det}(\mathbf{I}_3+\bm{\Psi}\cdot\mathbf{b})]^{-\kappa}\;\exp\!\left[-\mathbf{b}:[(\mathbf{I}_3+\bm{\Psi}\cdot\mathbf{b})^{-1}\cdot\bm{\Psi}\cdot\bm{\Theta}]\right] \, .
\label{Eq_signal_Magic-DIAMOND}
\end{equation}
Using our results in Equations~\ref{Eq_first_moment} and \ref{Eq_second_moment}, we retrieved the mean diffusion tensor of the nc-mv-Gamma distribution,
\begin{equation}
\langle\mathbf{D}\rangle=\left.\frac{\partial\mathcal{M}_\Gamma}{\partial\mathbf{Z}}\right|_{\mathbf{Z}=\mathbf{0}}=\bm{\Psi}\cdot\left[\kappa\mathbf{I}_3+\bm{\Theta}\right] ,
\label{Eq_average_diffusion_tensor_Gamma}
\end{equation}
and computed its covariance tensor for the first time:
\begin{equation}
\mathbb{C}=\langle\mathbf{D}^{\otimes 2}\rangle-\langle\mathbf{D}\rangle^{\otimes\!\,2}= \left.\frac{\partial^2\mathcal{M}_\Gamma}{\partial^2\mathbf{Z}}\right|_{\mathbf{Z}=\mathbf{0}} - \left.\frac{\partial\mathcal{M}_\Gamma}{\partial\mathbf{Z}}\right|_{\mathbf{Z}=\mathbf{0}}^{\otimes 2} =\kappa\,\bm{\Psi}\otimes\bm{\Psi}+(\bm{\Psi}\cdot\bm{\Theta})\otimes\bm{\Psi}+\bm{\Psi}\otimes(\bm{\Theta}\cdot\bm{\Psi}) \, .
\label{Eq_covariance_matrix_Gamma}
\end{equation}
The proofs underlying these results are available in Appendix~\ref{Sec_proof}. These expressions enable the direct computation of the statistical descriptors in Section~\ref{Sec_diffusion_metrics} for the non-central matrix-variate Gamma distribution according to Equation~\ref{Eq_statistical_descriptors_D_C}.

\subsubsection{The matrix-variate Gamma approximation}
\label{Sec_mv_Gamma_approximation}

A new dMRI signal representation, dubbed "matrix-variate Gamma approximation", can be attained by describing the intra-voxel diffusion profile with a single nc-mv-Gamma distribution (see Equation~\ref{Eq_nc_mv_Gamma_dist}). To facilitate the implementation of this approximation, one first notices that $\bm{\Psi}$ and $\bm{\Theta}$ commute according to Equation~\ref{Eq_average_diffusion_tensor_Gamma}, because these tensors are both symmetric and their product, $\bm{\Psi}\cdot\bm{\Theta}=\langle\mathbf{D}\rangle-\kappa\bm{\Psi}$, is also symmetric. This commutation ensures that the covariance tensor in Equation~\ref{Eq_covariance_matrix_Gamma} is symmetric. Besides, the commutation of the Hermitian tensors $\bm{\Psi}$ and $\bm{\Theta}$ (here real symmetric) implies that they share the same eigenvectors, which also coincide with the eigenvectors of $\langle\mathbf{D}\rangle$ \textit{via} Equation~\ref{Eq_average_diffusion_tensor_Gamma}. Consequently, the covariance tensor Equation~\ref{Eq_covariance_matrix_Gamma}, of size $6\times 6$ in Mandel notation, can only be non-zero within its upper-left $3\times 3$ block (see Section~\ref{Sec_diffusion_metrics}), thereby making this covariance tensor effectively $3\times 3$.

Another ease of implementation can be provided by replacing the symmetric tensor $\bm{\Theta}$ (featuring unbounded eigenvalues) by the symmetric positive definite tensor 
\begin{equation}
\mathbf{H}^{-1} = \kappa\mathbf{I}_3+\bm{\Theta}\in\mathrm{Sym}^+(3) \, ,
\label{Eq_H}
\end{equation}
which features strictly positive eigenvalues. The choice to denote this tensor by $\mathbf{H}^{-1}$ comes form the fact that $1/\kappa$ relates to the "width" (heterogeneity) of the nc-mv-Gamma distribution (in the same way that the inverse of the concentration parameter of a Watson distribution relates to dispersion~\citep{Zhang_NODDI:2012}), and that $\bm{\Theta}$ has already been used to provide an additional shape parameter $\kappa^\prime$ in Refs.~\onlinecite{Scherrer_aDIAMOND:2017, Reymbaut_arxiv_Magic_DIAMOND:2020}. Note that $\mathbf{H}$ and $\mathbf{H}^{-1}$ commute with $\langle\mathbf{D}\rangle$ and $\bm{\Psi}$, so that they share the same eigenvectors. 

Rewriting Equations~\ref{Eq_signal_Magic-DIAMOND} and \ref{Eq_covariance_matrix_Gamma} as a function of $\langle\mathbf{D}\rangle$, $\mathbf{H}$ and $\bm{\Psi}=\langle\mathbf{D}\rangle\cdot\mathbf{H}$, one obtains the parametric diffusion signal
\begin{equation}
\mathcal{S}_\Gamma(\mathbf{b}) = \mathcal{S}_0 \,[\mathrm{Det}(\mathbf{I}_3+\bm{\Psi}\cdot\mathbf{b})]^{-\kappa}\;\mathrm{exp}\!\left[-\mathbf{b}:[(\mathbf{I}_3+\bm{\Psi}\cdot\mathbf{b})^{-1}\cdot\bm{\Psi}\cdot(\mathbf{H}^{-1}-\kappa\mathbf{I}_3)]\right]
\label{Eq_new_signal}
\end{equation}
and the covariance tensor
\begin{equation}
\mathbb{C}=\langle\mathbf{D}\rangle\otimes\bm{\Psi}+\bm{\Psi}\otimes\langle\mathbf{D}\rangle-\kappa\,\bm{\Psi}\otimes\bm{\Psi} \, .
\label{Eq_new_covariance}
\end{equation}
Fitting $\mathcal{S}_\Gamma(\mathbf{b})$ in Equation~\ref{Eq_new_signal} requires 11 parameters: $\mathcal{S}_0$, $\kappa$, the three eigenvalues of $\bm{\Psi}$, the three eigenvalues of $\mathbf{H}$, and the three Euler angles giving the orientation of the eigenvectors shared by $\bm{\Psi}$ and $\mathbf{H}$. If numerically non positive semidefinite, $\mathbb{C}$ can be substituted with the nearest symmetric positive semidefinite tensor ("nearest" in terms of the Frobenius norm) using Ref.~\onlinecite{Higham:1988}.


\end{widetext}

\section{Methods}
\label{Sec_Methods}

\subsection{In vivo human-brain data}
\label{Sec_in_vivo_data}

\begin{figure}[ht!]
\begin{center}
\includegraphics[width=20.5pc]{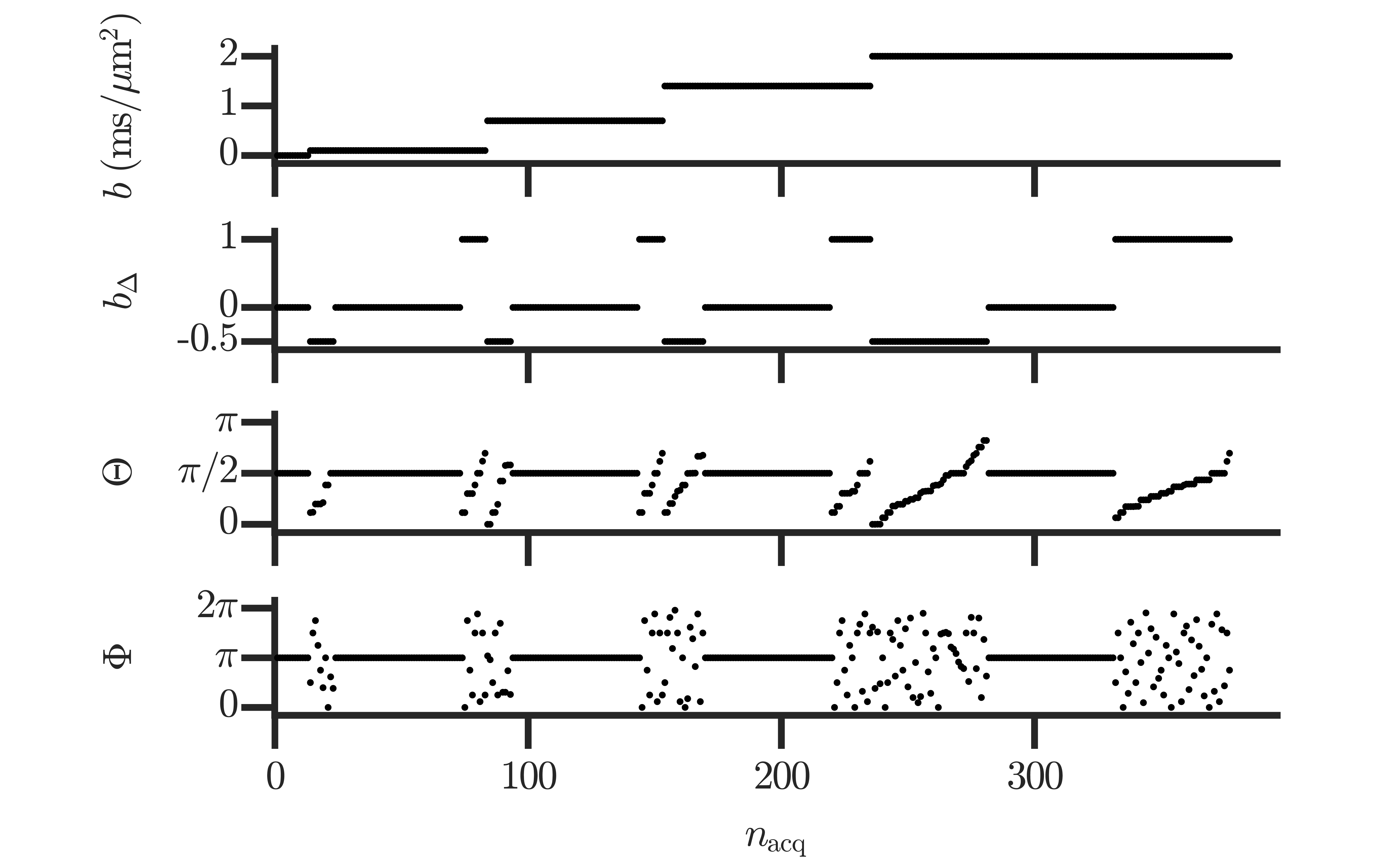}
\caption{Visualization of the acquisition scheme retrieved from Ref.~\onlinecite{Szczepankiewicz_data:2019} and used in this present work. Acquisition parameters, \textit{i.e.} the trace (size) $b$, normalized anisotropy (shape) $b_\Delta \in [-0.5, 1]$ and orientation $(\Theta,\Phi)$ of the b-tensor, are plotted as a function of sorted acquisition point index $n_\mathrm{acq}$. In particular, $b_\Delta = -0.5$, 0 and 1 correspond to planar, spherical and linear b-tensors, respectively.}
\label{Figure_acq}
\end{center}
\end{figure}

After implementing the matrix-variate Gamma approximation of Section~\ref{Sec_mv_Gamma_approximation} in Matlab, extending the framework of Refs.~\onlinecite{Matlab_toolbox,Nilsson_ISMRM:2018}, we evaluated it on a healthy human-brain `tensor-valued' dMRI dataset readily available online.~\citep{Szczepankiewicz_data:2019} This comprehensive dataset was acquired on a MAGNETOM 3T Prisma (Siemens Healthcare, Germany) using a prototype spin-echo sequence customized to support tensor-valued diffusion encoding,~\citep{Szczepankiewicz_DIVIDE:2019} and an echo-planar imaging (EPI) readout.~\citep{Mansfield_EPI:1977,Ordidge_EPI:1981} Its imaging parameters are $\mathrm{TR}=3.2 \;\mathrm{s}$, $\mathrm{TE}=91\;\mathrm{ms}$, $\mathrm{FOV}=220\times 220\times 60 \;\mathrm{mm}^3$, $\mathrm{matrix}=92\times 92\times 25$, $\mathrm{resolution}=2.5\times 2.5\times 2.5 \;\mathrm{mm}^3$, partial-$\mathrm{Fourier}=7/8$, $\mathrm{bandwidth}=1940 \;\mathrm{Hz/pix}$, and echo $\mathrm{spacing}=0.6\; \mathrm{ms}$. The sequence used interleaved slice excitation, strong fat saturation, in-plane acceleration $\mathrm{iPAT}=2$ with GRAPPA reconstruction and 30 reference lines. Tensor-valued diffusion encoding was performed with numerically optimized,~\citep{Sjolund:2015} Maxwell-compensated,~\citep{Szczepankiewicz_Maxwell:2019} waveforms. The spectral tuning of these waveforms~\citep{Lundell:2019} is discussed in Ref.~\onlinecite{Szczepankiewicz_data:2019}. The dataset was motion- and eddy-corrected by registering the images to an extrapolated reference~\citep{Nilsson:2015} using Elastix.~\citep{Klein_Elastix:2010} Its signal-to-noise ratio (SNR) was estimated to 30 in the corona radiata using the spherically encoded diffusion signal at $b=0.1\;\mathrm{ms}/\text{\textmu}\mathrm{m}^2$ (see Supplemental Material of Ref.~\onlinecite{Szczepankiewicz_DIVIDE:2019}). The overall acquisition scheme is illustrated in Figure~\ref{Figure_acq}.

\subsection{In silico data}
\label{Sec_in_silico_data}

We evaluated the matrix-variate Gamma approximation \textit{in silico}, and compared it with the covariance tensor approximation~\citep{Westin:2016} on the basis of estimating the statistical descriptors of interest presented in Section~\ref{Sec_diffusion_metrics}, \textit{i.e} the mean isotropic diffusivity $\mathrm{E}[\mathit{D}_\mathrm{iso}]$, the normalized mean squared anisotropy $\tilde{\mathrm{E}}[\mathit{D}_\mathrm{aniso}^2]=\mathrm{E}[\mathit{D}_\mathrm{aniso}^2]/\mathrm{E}[\mathit{D}_\mathrm{iso}]^2$ and the variance of isotropic diffusivities $\mathrm{V}[\mathit{D}_\mathrm{iso}]$. To that end, we used the same process as that found in Ref.~\onlinecite{Reymbaut_accuracy_precision:2020}:
\begin{enumerate}
\item A system of interest is simulated by generating a set of ground-truth features $\{\mathit{D}_\parallel,\mathit{D}_\perp, \theta, \phi, \mathit{f}\}$ (with the signal fraction $f$ of a given diffusion component), from which the ground-truth statistical descriptors of interest and the ground-truth set of signals $\{\mathcal{S}_{\mathrm{gt},\mathit{i}}\}$, with acquisition index $i$, are computed using the acquisition scheme of Section~\ref{Sec_in_vivo_data} and a discretized version of Equation~\ref{Eq_signal_tensor_distribution}.
\item Each signal representation is run on an identical set of signals with added Rician noise: $\{\mathcal{S}_{\mathit{i}}\} = \{[(\mathcal{S}_{\mathrm{gt},\mathit{i}} + \nu_{\mathit{i}}/\mathrm{SNR})^2 + (\nu^\prime_{\mathit{i}}/\mathrm{SNR})^2]^{1/2}\}$, where $\nu_{\mathit{i}}$ and $\nu^\prime_{\mathit{i}}$ denote random numbers drawn from a normal distribution with zero mean and unit standard deviation. 
\item Step 2 is repeated 100 times to build up statistics on parameter estimation of $\mathrm{E}[\mathit{D}_\mathrm{iso}]$, $\tilde{\mathrm{E}}[\mathit{D}_\mathrm{aniso}^2]$ and $\mathrm{V}[\mathit{D}_\mathrm{iso}]$.
\end{enumerate}
In this work, we investigated a clinically relevant SNR of 30 and the ideal infinite SNR. Since the Rician bias has been shown to be relevant only when the SNR approaches five,\cite{Gudbjartsson_Patz:1995} affecting the estimation of diffusion metrics,\cite{Jones_Basser:2004,Gilbert:2007,Sotiropoulos:2013} the results presented in Section~\ref{Sec_Results} are identical to those obtained from signals with added Gaussian noise. While accuracy is quantified for a given estimation by the difference between the median estimation across noise realizations and the ground-truth (bias), precision is quantified by the interquartile range of estimations across noise realizations. A "good" estimation is defined as one that presents both high accuracy and high precision.

\newpage

\section{Results}
\label{Sec_Results}

The results presented in this section and discussed in Section~\ref{Sec_Discussion} aim to evaluate the matrix-variate Gamma approximation of Section~\ref{Sec_mv_Gamma_approximation} \textit{in vivo} and \textit{in silico}, thereby offering a proof of concept of the applicability of the matrix moments derived in Section~\ref{Sec_matrix_moments_D_C}. Alternatively, these results allow to identify the nature of the diffusion information that can be captured by non-central matrix-variate Gamma distributions (as those used in the DIAMOND model~\citep{Scherrer_DIAMOND:2016, Scherrer_aDIAMOND:2017, Reymbaut_arxiv_Magic_DIAMOND:2020}).

Figure~\ref{Figure_signals} presents the fitted signals yielded by the matrix-variate Gamma approximation \textit{in vivo}, within typical voxels associated with cerebrospinal fluid (CSF), grey matter (GM), single-fiber white matter (WM) in the corpus callosum, and crossing fibers in the anterior centrum semiovale. These fitted signals are compared with the signals measured using the acquisition scheme detailed in Section~\ref{Sec_in_vivo_data}. Figure~\ref{Figure_maps} shows axial parameter maps of the statistical descriptors in Section~\ref{Sec_diffusion_metrics} estimated by the matrix-variate Gamma approximation. Unexpected discrepancies between the results of Figure~\ref{Figure_maps} and the known anatomy are further investigated \textit{in silico} in Figures~\ref{Figure_in_silico_iso}, \ref{Figure_in_silico_aniso} and \ref{Figure_in_silico_mixed}, which feature the statistical descriptors estimated with the matrix-variate Gamma approximation and the covariance tensor approximation of Ref.~\onlinecite{Westin:2016} in numerical systems consisting of isotropic components only, anisotropic components only, and a mixture of isotropic and anisotropic components, respectively. This \textit{in silico} study followed the process detailed in Section~\ref{Sec_in_silico_data}.

\section{Discussion}
\label{Sec_Discussion}

In this section, let us denote the matrix-variate Gamma approximation by "mv-Gamma" and the covariance tensor approximation by "Cov" for compactness.

\begin{figure*}[ht!]
\begin{center}
\includegraphics[width=33pc]{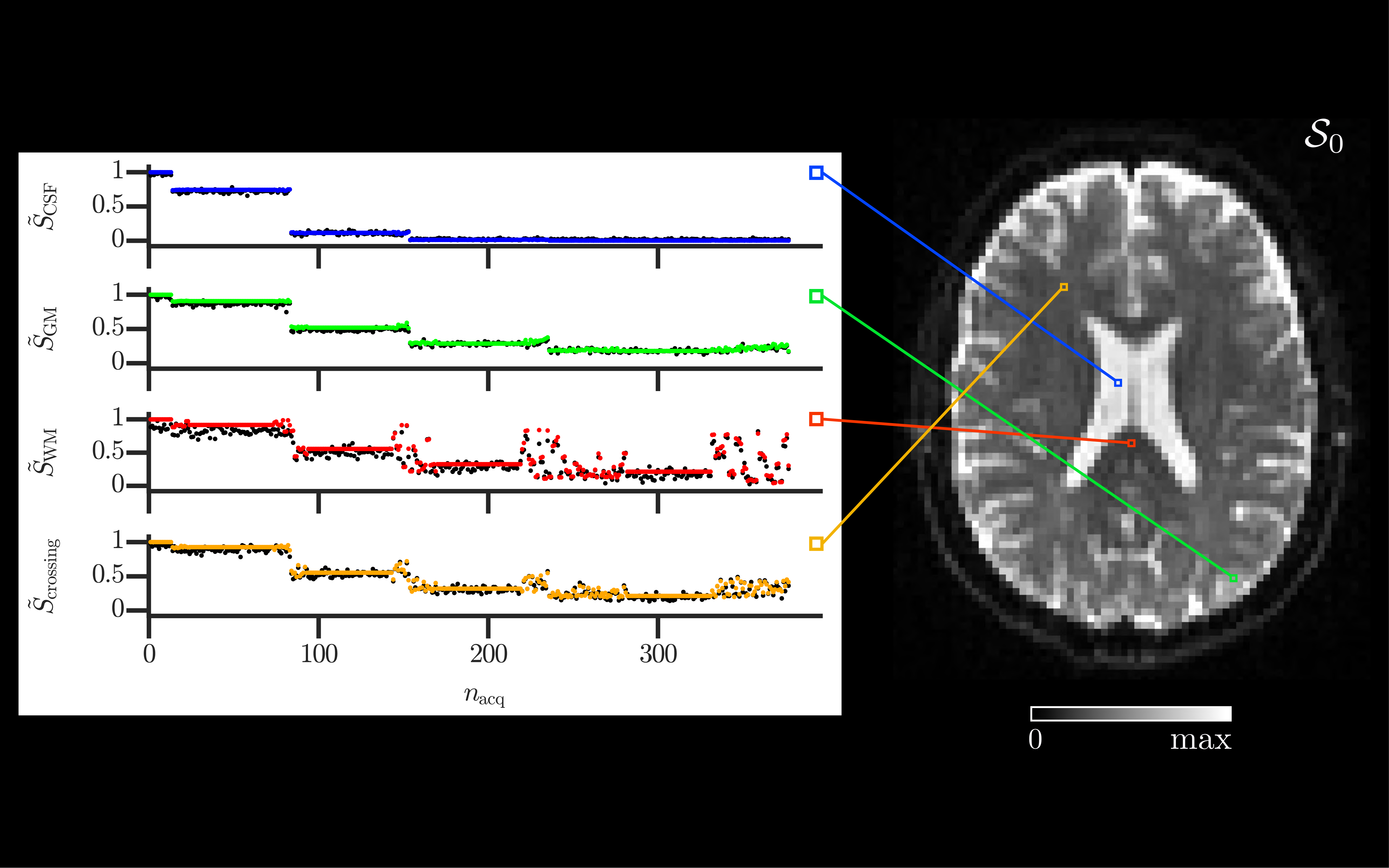}
\caption{Signals fitted within the mv-Gamma for archetypal voxel contents. Normalized signals $\tilde{S} = \mathcal{S}/\mathcal{S}_0$ are plotted as a function of the sorted acquisition point index $n_\mathrm{acq}$ of Figure~\ref{Figure_acq}. While black dots correspond to the measured normalized signals, colored dots are associated with the fitted normalized signals in four voxels of interest illustrated on a $\mathcal{S}_0$ map: a blue voxel in the cerebrospinal fluid (CSF) of a ventricle, a green voxel in cortical grey matter (GM), a red voxel in the white matter (WM) of the corpus callosum, and an orange voxel in an area of crossing within the anterior centrum semiovale.}
\label{Figure_signals}
\end{center}
\end{figure*}

\begin{figure*}[ht!]
\begin{center}
\includegraphics[width=33pc]{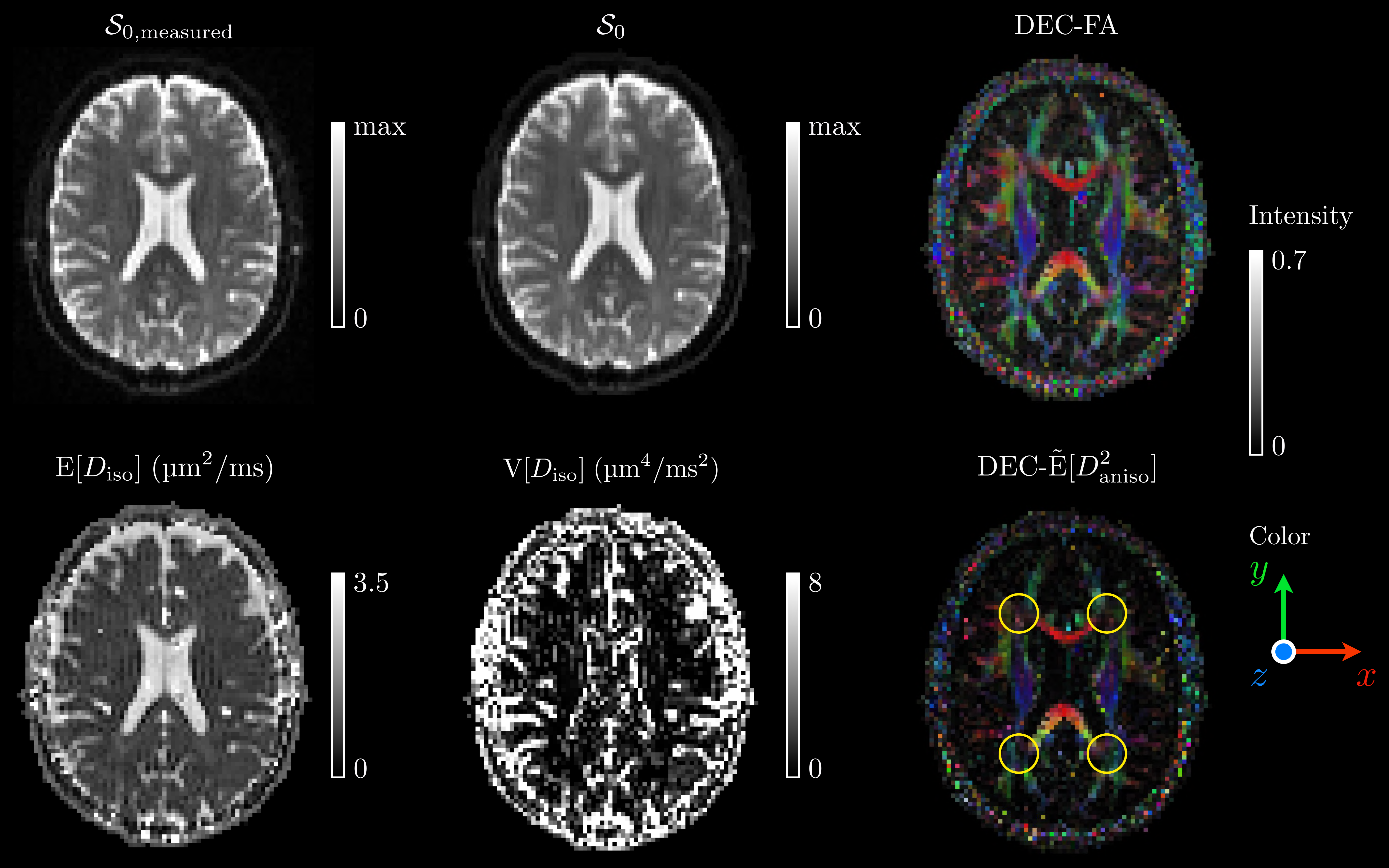}
\caption{Parameter maps estimated with mv-Gamma. The greyscale maps feature the average measured $b=0$ signal $\mathcal{S}_{0,\text{measured}}$ for comparison with the estimated $\mathcal{S}_0$ map, and the mean isotropic diffusivity $\mathrm{E}[D_\mathrm{iso}]$ and variance of isotropic diffusivities $\mathrm{V}[D_\mathrm{iso}]$ (see Section~\ref{Sec_diffusion_metrics}). In particular, $\mathrm{V}[D_\mathrm{iso}]$ presents large values compared to a reasonable upper bound of $1.21\;\text{\textmu}\mathrm{m}^4/\mathrm{ms}^2$ obtained for a voxel containing equal proportions of free water ($D_\mathrm{iso} = 3\;\text{\textmu}\mathrm{m}^2/\mathrm{ms}$) and white matter ($D_\mathrm{iso} = 0.8\;\text{\textmu}\mathrm{m}^2/\mathrm{ms}$). As for the directionally encoded color (DEC) maps,~\citep{Pajevic_Pierpaoli:1999} their respective intensities are given by the fractional anisotropy (FA)~\citep{Basser_Pierpaoli:1996} and the normalized mean squared anisotropy $\tilde{\mathrm{E}}[\mathit{D}_\mathrm{aniso}^2]$. Their colors code for the orientation of the main eigenvector $\mathbf{u}\equiv (u_x,u_y, u_z)$ of the mean diffusion tensor $\langle \mathbf{D}\rangle$ retrieved with mv-Gamma (see Equation~\ref{Eq_average_diffusion_tensor_Gamma}) according to $[\mathrm{red},\mathrm{green},\mathrm{blue}] = [\vert u_x\vert, \vert u_y\vert, \vert u_z\vert]$, where $x$, $y$ and $z$ correspond to the "left-right", "anterior-posterior" and "superior-inferior" directions, respectively. The yellow circles highlight regions of the centrum semiovale where the normalized mean squared anisotropy $\tilde{\mathrm{E}}[\mathit{D}_\mathrm{aniso}^2]$ vanishes unexpectedly.}
\label{Figure_maps}
\end{center}
\end{figure*}

As seen from Figure~\ref{Figure_signals}, mv-Gamma fits the dMRI signal rather adequately in various brain regions, even in areas of fiber crossings. However, Figure~\ref{Figure_maps} indicates that even though this signal representation yields maps of $\mathcal{S}_0$, $\mathrm{E}[D_\mathrm{iso}]$, $\mathrm{V}[D_\mathrm{iso}]$ and directionally encoded color (DEC)~\citep{Pajevic_Pierpaoli:1999} fractional anisotropy (FA)~\citep{Basser_Pierpaoli:1996} that are consistent with the known anatomy, it also gives anomalously large values of $\mathrm{V}[D_\mathrm{iso}]$ and a DEC normalized mean squared anisotropy $\tilde{\mathrm{E}}[\mathit{D}_\mathrm{aniso}^2]$ that vanishes unexpectedly in fiber-crossing regions. Nevertheless, the $\mathrm{V}[D_\mathrm{iso}]$ map exhibits the contrast expected from $\mathrm{V}[D_\mathrm{iso}]$, \textit{i.e.} large values at the interface between WM and the CSF in the ventricles, and at the interface between cortical GM and the CSF surrounding the brain, and small values otherwise. 

\begin{figure*}[ht!]
\begin{center}
\includegraphics[width=33pc]{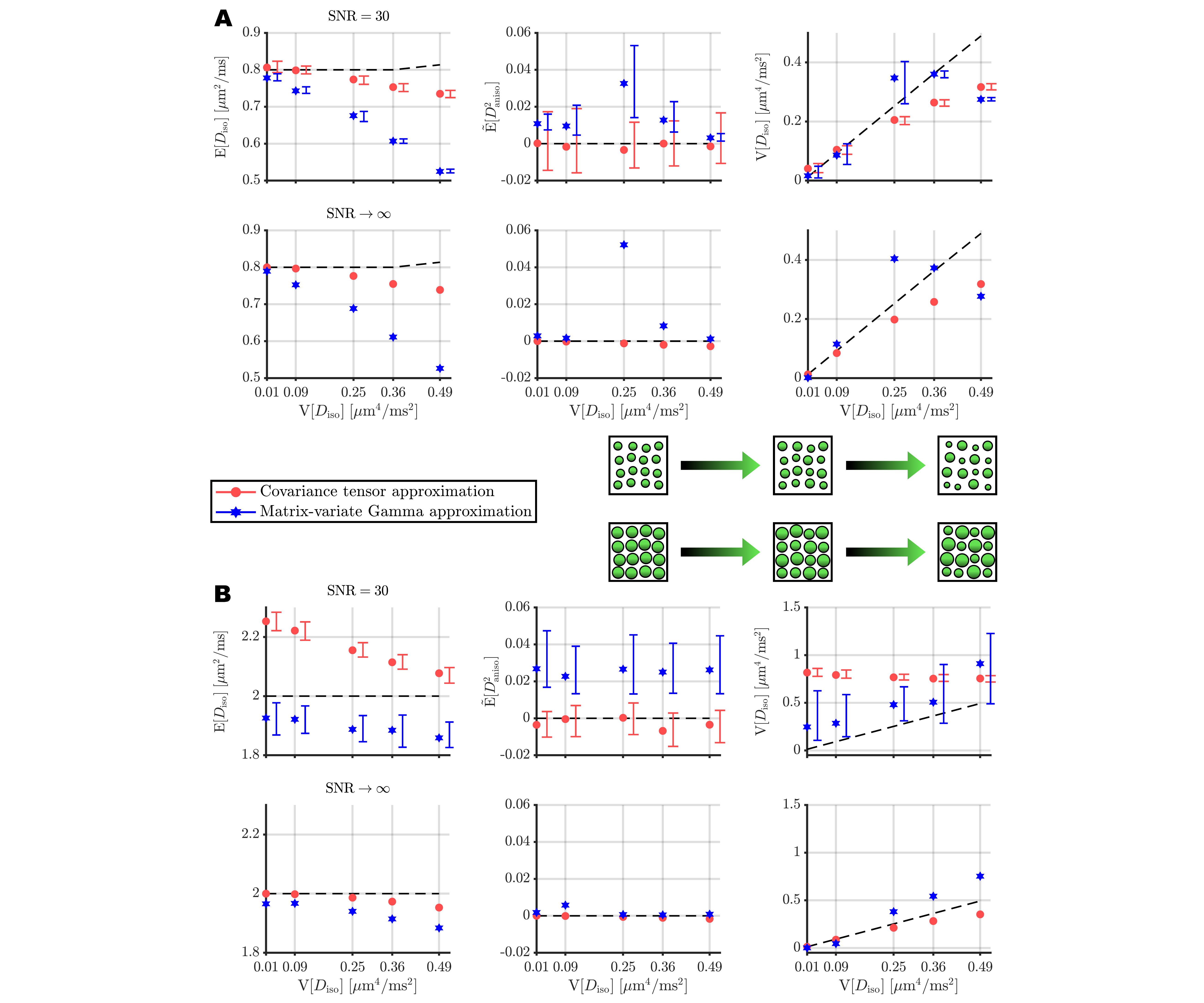}
\caption{Statistical descriptors of interest estimated by mv-Gamma and Cov for \textit{in silico} systems consisting of isotropic components described by a bimodal distribution of $D_\mathrm{iso}$ with constant mean isotropic diffusivity $\mathrm{E}[D_\mathrm{iso}]$ of (A) $0.8\;\text{\textmu}\mathrm{m}^2/\mathrm{ms}$ and (B) $2\;\text{\textmu}\mathrm{m}^2/\mathrm{ms}$. Each system was designed as a sum of two Gaussian distributions with identical standard deviations ($\sigma = 0.05\;\text{\textmu}\mathrm{m}^2/\mathrm{ms}$) whose separation sets the value of $\mathrm{V}[D_\mathrm{iso}]$. Within each of panels A and B, the properties of the simulated system - mean isotropic diffusivity $\mathrm{E}[D_\mathrm{iso}]$, normalized mean squared anisotropy $\tilde{\mathrm{E}}[D_\mathrm{aniso}^2]$, and variance of isotropic diffusivities $\mathrm{V}[D_\mathrm{iso}]$ - were estimated by fitting Cov (red circles) and mv-Gamma (blue stars) to 100 different noise realizations of the ground-truth signal at either finite (top) or infinite (bottom) SNR (see Section~\ref{Sec_in_silico_data}). The various descriptors are plotted as a function of $\mathrm{V}[D_\mathrm{iso}]$, with symbols indicating the medians of the descriptors across 100 noise realizations, shifted error bars representing their interquartile ranges across 100 noise realizations, and black dashed lines denoting the ground-truth descriptors. The center-right boxes illustrate the diffusion tensor distributions of the investigated voxel contents in terms of green glyphs.}
\label{Figure_in_silico_iso}
\end{center}
\end{figure*}

\begin{figure*}[ht!]
\begin{center}
\includegraphics[width=33pc]{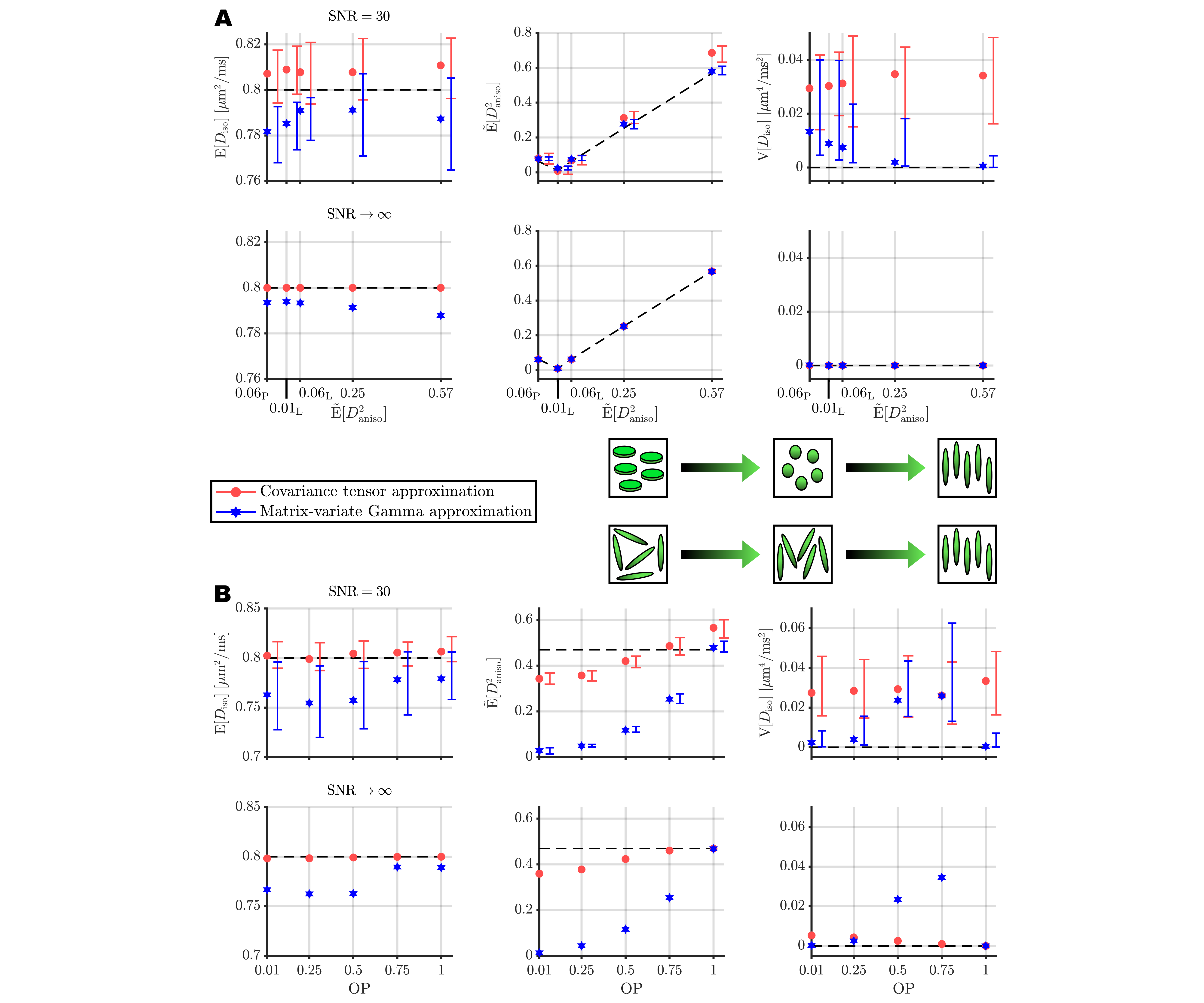}
\caption{Statistical descriptors of interest estimated by mv-Gamma and Cov for \textit{in silico} systems consisting of anisotropic components, each of which is described by a unimodal distribution of $D_\Delta$ (Gaussian distribution) with constant $D_\mathrm{iso} = 0.8\;\text{\textmu}\mathrm{m}^2/\mathrm{ms}$ and constant standard-deviation-to-the-mean ratio of 0.1. (A) Coherent (orientationally ordered) systems with varying normalized mean squared anisotropy $\tilde{\mathrm{E}}[D_\mathrm{aniso}^2]$. Notice that even though the first and third simulated systems exhibit the same normalized mean squared anisotropy, the former contains planar anisotropic components and the latter contains linear anisotropic components, as indicated by the subscripts "P" and "L" on the $\tilde{\mathrm{E}}[D_\mathrm{aniso}^2]$ axis, respectively. (B) Systems with varying orientational order parameter $\mathrm{OP}=\mathrm{E}[P_2(\cos\beta)]$, where $P_2(x) = (3x^2-1)/2$ is the second Legendre polynomial and $\beta$ denotes the shortest angle between a component $(\theta,\phi)$ and the main eigenvector of the voxel-scale Saupe order tensor.~\citep{Lasic:2014,Topgaard_liquid:2016} Symbol/color conventions are identical to those of Figure~\ref{Figure_in_silico_iso}.}
\label{Figure_in_silico_aniso}
\end{center}
\end{figure*}

\begin{figure*}[ht!]
\begin{center}
\includegraphics[width=33pc]{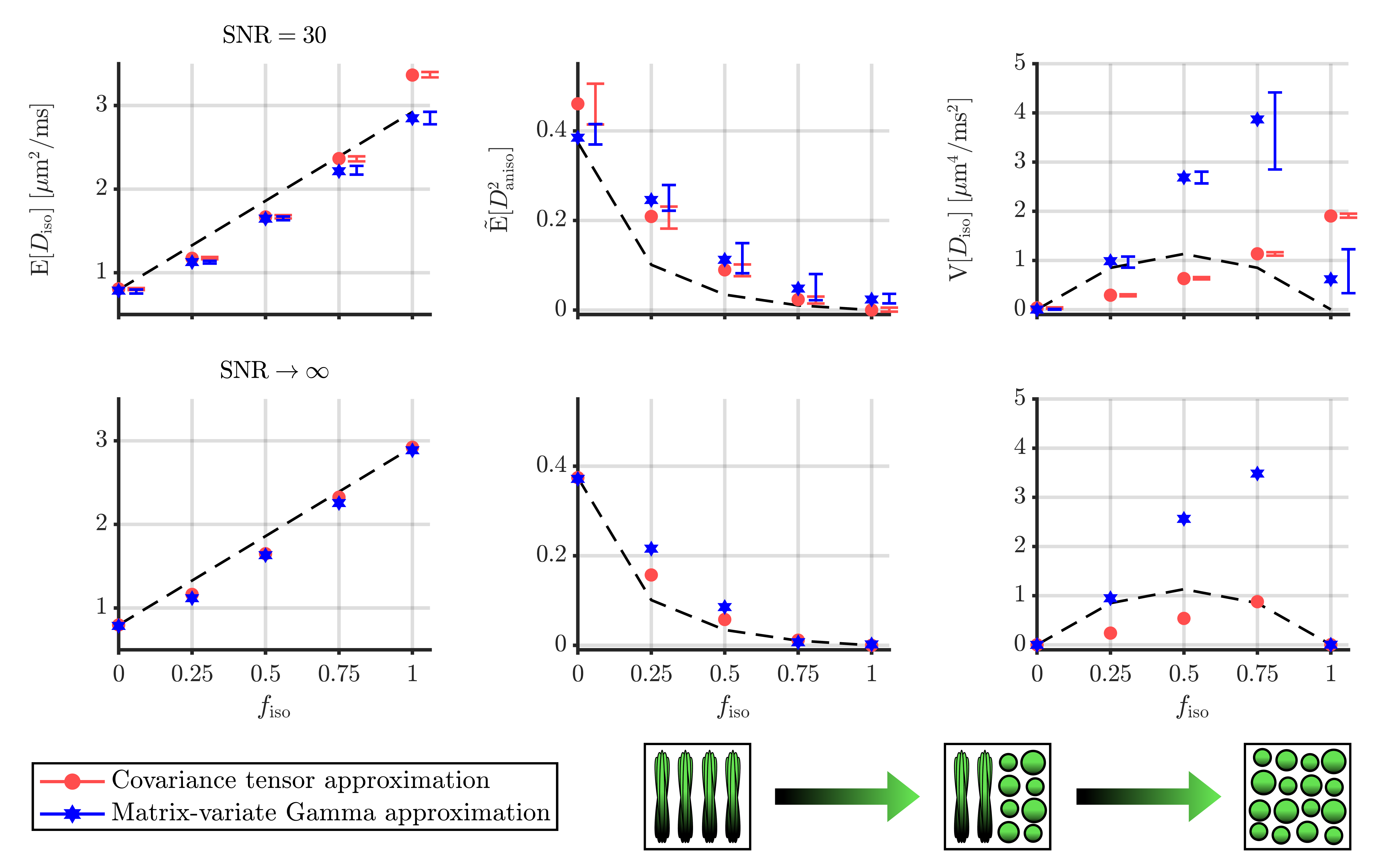}
\caption{Statistical descriptors of interest estimated by mv-Gamma and Cov for \textit{in silico} systems consisting of a mixture of an isotropic CSF-like structure (total signal fraction $\mathit{f}_{\mathrm{iso}}\in [0,1]$, Gaussian distribution of $\mathit{D}_{\mathrm{iso}}$ with mean $3$~{\textmu}m$^2$/ms and standard deviation $0.1$~{\textmu}m$^2$/ms) and an anisotropic structure (total signal fraction equal to $1-\mathit{f}_{\mathrm{iso}}$). The anisotropic structure is made of a Watson distribution of anisotropic components with an internal distribution of $\mathit{D}_\parallel$ (Gaussian distribution centered at $1.77$~{\textmu}m$^2$/ms) and $\mathit{D}_\perp$ (Gaussian distribution centered at $0.31$~{\textmu}m$^2$/ms). These diffusivities come from typical values estimated by the Magic DIAMOND model.~\cite{Reymbaut_arxiv_Magic_DIAMOND:2020} The Watson distribution is set to an intermediate orientational order parameter $\mathrm{OP}=0.4$ to mimic some fiber dispersion. Symbol/color conventions are identical to those of Figure~\ref{Figure_in_silico_iso}.}
\label{Figure_in_silico_mixed}
\end{center}
\end{figure*}

These trends are confirmed \textit{in silico} in Figures~\ref{Figure_in_silico_iso}, \ref{Figure_in_silico_aniso} and \ref{Figure_in_silico_mixed}, which we first discuss. In general, the estimations of mv-Gamma and Cov share similar precision, except when estimating $\mathrm{V}[D_\mathrm{iso}]$ at high $\mathrm{E}[D_\mathrm{iso}]$ (see Figures~\ref{Figure_in_silico_iso} and \ref{Figure_in_silico_mixed}), where Cov is more precise. In terms of accuracy, both signal representations share similar infinite-SNR biases in nature, although generally more pronounced for mv-Gamma. These biases persist at finite SNR, with mv-Gamma being more accurate than Cov in estimating $\mathrm{E}[D_\mathrm{iso}]$ in systems with low $\mathrm{V}[D_\mathrm{iso}]$ (see Figures~\ref{Figure_in_silico_iso}.B and \ref{Figure_in_silico_mixed}), $\mathrm{V}[D_\mathrm{iso}]$ in systems with high $\mathrm{E}[D_\mathrm{iso}]$ (see Figures~\ref{Figure_in_silico_iso}.B and \ref{Figure_in_silico_mixed}), and $\mathrm{V}[D_\mathrm{iso}]$ and $\tilde{\mathrm{E}}[\mathit{D}_\mathrm{aniso}^2]$ in coherent (orientationally ordered) anisotropic systems (see Figure~\ref{Figure_in_silico_aniso}.A). In particular, mv-Gamma performs better than Cov in coherent anisotropic systems of high prolate anisotropy. However, while Figure~\ref{Figure_in_silico_aniso}.B demonstrates that the normalized mean squared anisotropy $\tilde{\mathrm{E}}[\mathit{D}_\mathrm{aniso}^2]$ estimated by mv-Gamma vanishes upon reducing the orientational order parameter~\citep{Lasic:2014,Topgaard_liquid:2016} of anisotropic systems (to a greater extent compared to Cov), Figure~\ref{Figure_in_silico_mixed} shows that the variance of isotropic diffusivities $\mathrm{V}[D_\mathrm{iso}]$ estimated by mv-Gamma can largely overshoot the ground-truth value in mixed systems with high $\mathrm{V}[D_\mathrm{iso}]$, thereby confirming the trends seen \textit{in vivo} in Figure~\ref{Figure_maps}.

These limitations of the matrix-variate Gamma approximation can be understood from a mathematical standpoint, in turn informing on the fundamental limitations of the nc-mv-Gamma distribution with regard to capturing intra-voxel heterogeneity. We showed in Section~\ref{Sec_mv_Gamma_approximation} that the covariance tensor of the nc-mv-Gamma distribution (see Equations~\ref{Eq_covariance_matrix_Gamma} and \ref{Eq_new_covariance}) can only be non-zero within its upper-left $3\times 3$ block in Mandel notation (see Section~\ref{Sec_diffusion_metrics}). However, the other $3\times 3$ blocks of the covariance tensor, zero in the case of the nc-mv-Gamma distribution, are mostly involved in capturing the heterogeneity of orientationally dispersed voxel contents, and the heterogeneity of voxel contents simultaneously featuring variances in the  trace (size) and anisotropy (shape) of the underlying diffusion tensors, as shown in Refs.~\onlinecite{Westin:2016,Magdoom:2020}. This explains the limitations of the matrix-variate Gamma approximation observed in Figures~\ref{Figure_maps}, \ref{Figure_in_silico_iso}, \ref{Figure_in_silico_aniso} and \ref{Figure_in_silico_mixed}.

\section{Conclusions}
\label{Sec_Conclusions}

We established practical mathematical tools, the matrix moments of the diffusion tensor distribution (DTD) $\mathcal{P}(\mathbf{D})$, that remove the fundamental barrier preventing a more widespread use of matrix-variate distributions as plausible approximations of the DTD: their intractability. Indeed, the matrix moments enable the computation of the mean diffusion tensor and covariance tensor of any matrix-variate parametric functional form chosen for the DTD, given that its moment-generating function is known. In turn, statistical descriptors of the DTD that are common to various methods can be extracted from these tensors, allowing to investigate the performance of a given functional approximation in capturing intra-voxel heterogeneity, and to compare multiple signal representations and models on the basis of estimating identical sets of descriptors. 

As a proof of concept, we computed the matrix moments of the non-central matrix-variate Gamma distribution, deriving its covariance tensor for the first time. Building upon these calculations, we designed a new signal representation wherein the intra-voxel diffusion profile is described by a single non-central matrix-variate Gamma distribution: the matrix-variate Gamma approximation. This approximation fails to capture the heterogeneity arising from orientation dispersion and from simultaneous variances in the size and shape of the underlying diffusion tensors, which can be understood from the structure of the covariance tensor associated with that specific choice of parametric distribution for $\mathcal{P}(\mathbf{D})$. However, the matrix-variate Gamma approximation performs well in orientationally ordered anisotropic systems of high prolate anisotropy, which justifies its use to describe anisotropic diffusion compartments, such as within the DIAMOND model.~\citep{Scherrer_DIAMOND:2016,Scherrer_aDIAMOND:2017, Reymbaut_arxiv_Magic_DIAMOND:2020} Finally, the matrix moments of the non-central matrix-variate Gamma distribution provide the DIAMOND model with fiber-specific statistical descriptors that are straightforwardly comparable to those of other fiber-specific techniques.~\citep{Assaf:2004, Assaf_CHARMED:2005, Reymbaut_arxiv_MC_DPC:2020} We encourage the diffusion MRI community to use these tools to experiment with other matrix-variate distributions~\citep{Gupta_Nagar_Book:2000} and to identify their respective advantages/limitations in capturing microstructural heterogeneity.

\section*{Acknowledgments}
This work was financially supported by the Swedish Foundation for Strategic Research (ITM17-0267) and the Swedish Research Council (2018-03697).

\newpage

\newpage

\newpage

\newpage

\clearpage
\newpage

\begin{appendix}

\begin{widetext}

\section{Validity of the DTD description}
\label{Sec_validity_DTD}

Let us discuss the validity of the DTD description found in Ref.~\onlinecite{Jian:2007} and formulated in Equation~\ref{Eq_signal_tensor_distribution}. This description is equivalent to considering a `snapshot' of the combined non-Gaussian diffusion effects of restriction and exchange at a given observational time-scale, and approximating the signal decay as a continuous weighted sum of exponential decays. As the dMRI observational time-scale depends on the spectral content of the diffusion-encoding gradients,~\citep{Stepisnik:1981, Stepisnik:1985, Callaghan_Stepisnik:1995} so do the set of exponential decays estimated in Equation~\ref{Eq_signal_tensor_distribution}, and the measured DTD $\mathcal{P}(\mathbf{D})$. Such time-dependent effects arise as a result of restricted diffusion~\citep{Woessner:1963} and exchange,~\citep{Johnson:1993,Li_Springer:2019} and have been measured in human-brain white matter,~\citep{Van:2014,Baron_Beaulieu:2014,Baron_Beaulieu:2015,Fieremans:2016,Veraart:2019,Lundell:2019,dellAcqua_ISMRM:2019} spinal cord~\citep{Jespersen:2018,Grussu:2019} and prostate~\citep{Lemberskiy:2017,Lemberskiy:2018} using pulse sequences specifically designed for varying the dMRI observational time-scale over extended ranges. Nevertheless, the DTD description holds for the limited range of long diffusion times probed by clinical dMRI experiments in the brain.~\citep{Clark:2001,Ronen:2006,Nilsson:2009, Nilsson:2013a, Nilsson:2013b, deSantis_T1:2016, Lampinen:2017,Veraart:2018,Grussu:2019,Szczepankiewicz_ISMRM:2019} 

\section{Validation on the matrix-variate Gaussian distribution}
\label{Sec_validation}

\setcounter{equation}{0}

The expressions Equations~\ref{Eq_first_moment} and \ref{Eq_second_moment} for the mean diffusion tensor $\langle \mathbf{D}\rangle$ and covariance tensor $\mathbb{C}$ of $\mathcal{P}(\mathbf{D})$ are validated below by using them to retrieve the known mean tensor $\mathbf{M}$ and covariance tensor $\bm{\Sigma}\otimes \bm{\Psi}$ of the mv-Gaussian distribution~\citep{Gupta_Nagar_Book:2000,Kollo_von_Rosen_book:2006} featured in Refs.~\onlinecite{Basser_Pajevic:2003, Pajevic_Basser:2003,Westin:2016}:
\begin{equation}
\mathcal{P}_\mathrm{Gauss}(\mathbf{X}) = \frac{1}{(2\pi)^{9/2}\, \mathrm{Det}(\bm{\Sigma})^{3/2} \mathrm{Det}(\bm{\Psi})^{3/2}}\, \exp\!\left[ \mathrm{Tr}\!\left( -\frac{1}{2}\, \bm{\Sigma}^{-1}\cdot (\mathbf{X}-\mathbf{M})\cdot \bm{\Psi}^{-1}\cdot (\mathbf{X}-\mathbf{M})^\mathrm{T} \right) \right] ,
\end{equation}
where $\mathbf{X}$ and $\mathbf{M}$ are arbitrary $3\times 3$ real matrices, and $\bm{\Sigma},\bm{\Psi}\in \mathrm{Sym}^+(3)$. To do so, we consider the moment-generating function of this distribution:~\citep{Gupta_Nagar_Book:2000}
\begin{equation}
\mathcal{M}_\mathrm{Gauss}(\mathbf{Z}) = \exp\!\left[ \mathrm{Tr}\!\left( \mathbf{Z}\cdot\mathbf{M} +\frac{1}{2}\, \mathbf{Z}^\mathrm{T}\cdot \bm{\Sigma} \cdot \mathbf{Z} \cdot \bm{\Psi} \right) \right] .
\end{equation}
Note that similar proofs can be found in Ref.~\onlinecite{Kollo_von_Rosen_book:2006}.

\subsection{Mean diffusion tensor}

Using Ref.~\onlinecite{Laue_website} to compute the first-order matrix derivative of $\mathcal{M}_\mathrm{Gauss}(\mathbf{Z})$, and transposing its result according to Section~\ref{Sec_matrix_calculus}, one has
\begin{equation}
\frac{\partial\mathcal{M}_\mathrm{Gauss}}{\partial\mathbf{Z}} = \exp\!\left[ \mathrm{Tr}\!\left( \mathbf{Z}\cdot\mathbf{M} +\frac{1}{2}\, \mathbf{Z}^\mathrm{T}\cdot \bm{\Sigma} \cdot \mathbf{Z} \cdot \bm{\Psi} \right) \right] \times \left[ \mathbf{M} + \bm{\Psi}\cdot\mathbf{Z}^\mathrm{T}\cdot \bm{\Sigma} \right] = \mathcal{M}_\mathrm{Gauss}(\mathbf{Z})\times \left[ \mathbf{M} + \bm{\Psi}\cdot\mathbf{Z}^\mathrm{T}\cdot \bm{\Sigma} \right] \, ,
\label{Eq_interm_20}
\end{equation}
which immediately gives 
\begin{equation}
\langle \mathbf{X}\rangle =  \left.\frac{\partial\mathcal{M}_\mathrm{Gauss}}{\partial\mathbf{Z}}\right\vert_{\mathbf{Z}=\mathbf{0}} = \mathbf{M}
\label{Eq_proof_M}
\end{equation}
using Equation~\ref{Eq_first_moment}.

\subsection{Covariance tensor}

The second-order matrix derivative of $\mathcal{M}_\mathrm{Gauss}(\mathbf{Z})$ equals the first-order matrix derivative of Equation~\ref{Eq_interm_20}. Using the scalar product rule of Equation~\ref{Eq_scalar_product_rule} and the fact that~\citep{Laue_website} $\partial (\bm{\Psi}\cdot\mathbf{Z}^\mathrm{T}\cdot \bm{\Sigma})/\partial \mathbf{Z} = \bm{\Sigma}\otimes \bm{\Psi}$, one obtains
\begin{equation}
\frac{\partial^2\mathcal{M}_\mathrm{Gauss}}{\partial\mathbf{Z}^2} = \frac{\partial\mathcal{M}_\mathrm{Gauss}}{\partial\mathbf{Z}}\otimes \left[ \mathbf{M} + \bm{\Psi}\cdot\mathbf{Z}^\mathrm{T}\cdot \bm{\Sigma} \right] + \mathcal{M}_\mathrm{Gauss}(\mathbf{Z})\times \bm{\Sigma} \otimes \bm{\Psi} \, ,
\end{equation}
which gives
\begin{equation}
\left.\frac{\partial^2\mathcal{M}_\mathrm{Gauss}}{\partial\mathbf{Z}^2}\right\vert_{\mathbf{Z}=\mathbf{0}} = \mathbf{M}^{\otimes 2} + \bm{\Sigma} \otimes \bm{\Psi}
\end{equation}
using Equation~\ref{Eq_proof_M}, so that
\begin{equation}
\mathbb{C} = \left.\frac{\partial^2\mathcal{M}_\mathrm{Gauss}}{\partial\mathbf{Z}^2}\right\vert_{\mathbf{Z}=\mathbf{0}} - \left.\frac{\partial\mathcal{M}_\mathrm{Gauss}}{\partial\mathbf{Z}}\right\vert_{\mathbf{Z}=\mathbf{0}}^{\otimes 2} = \bm{\Sigma} \otimes \bm{\Psi}
\end{equation}
using Equation~\ref{Eq_second_moment}.

\section{Proofs for the non-central matrix-variate Gamma distribution}
\label{Sec_proof}

\setcounter{equation}{0}

\subsection{Average diffusion tensor}

Starting from the moment-generating function $\mathcal{M}_\Gamma$ of the nc-mv-Gamma distribution in Equation~\ref{Eq_moment_generating_function_gamma}, one uses Ref.~\onlinecite{Laue_website}, transposing its result according to Section~\ref{Sec_matrix_calculus}, to obtain
\begin{align}
\frac{\partial \mathcal{M}_\Gamma}{\partial \mathbf{Z}} 
 & = \left[\mathrm{Det}(\mathbf{I}_3-\bm{\Psi}\cdot \mathbf{Z})\right]^{-\kappa} \exp\!\left[\mathrm{Tr}\!\left(\left[(\mathbf{I}_3-\bm{\Psi}\cdot \mathbf{Z})^{-1}-\mathbf{I}_3\right]\cdot\bm{\Theta}\right)\right] \nonumber \\
 & \quad \times \left\{\kappa \left[\mathrm{Det}(\mathbf{I}_3-\bm{\Psi}\cdot \mathbf{Z})\right]^{-1}\times \bm{\Psi}\cdot\left[\mathrm{Adj}(\mathbf{I}_3-\bm{\Psi}\cdot \mathbf{Z})\right]^\text{T}   + \bm{\Psi}\cdot \left[(\mathbf{I}_3-\bm{\Psi}\cdot \mathbf{Z})^{-1}\right]^\text{T}\cdot \bm{\Theta} \cdot \left[(\mathbf{I}_3-\bm{\Psi}\cdot \mathbf{Z})^{-1}\right]^\text{T} \right\}\,,
\label{Eq_first_derivative}
\end{align}
where the adjugate matrix (also called classical adjoint, or adjunct) is defined for a $n\times n$ matrix $\mathbf{A}$ by
\begin{equation}
\mathbf{A}\cdot\mathrm{Adj}(\mathbf{A})=\mathrm{Adj}(\mathbf{A})\cdot\mathbf{A} = \mathrm{Det}(\mathbf{A})\, \mathbf{I}_n\,.
\label{Eq_adj}
\end{equation}
Let us simplify the expression of Equation~\ref{Eq_first_derivative}. First, one has
\begin{equation}
\left[ \left(\mathbf{I}_3-\bm{\Psi}\cdot \mathbf{Z} \right)^{-1} \right]^\text{T} = \left[ \left(\mathbf{I}_3-\bm{\Psi}\cdot \mathbf{Z} \right)^\text{T} \right]^{-1} = \left(\mathbf{I}_3- \mathbf{Z}^\text{T}\cdot \bm{\Psi}^\text{T} \right)^{-1} = \left(\mathbf{I}_3- \mathbf{Z}\cdot \bm{\Psi} \right)^{-1}\,.
\label{Eq_interm_1}
\end{equation}
Second, Equation~\ref{Eq_interm_1} indicates that $\mathbf{I}_3-\bm{\Psi}\cdot \mathbf{Z}$ is invertible because of the condition on $\mathbf{Z}$ ensuring the convergence of the moment-generating function Equation~\ref{Eq_moment_generating_function_gamma}. This implies \textit{via} Equation~\ref{Eq_adj} that its adjugate satisfies
\begin{equation}
\mathrm{Adj}(\mathbf{I}_3-\bm{\Psi}\cdot \mathbf{Z}) = \mathrm{Det}(\mathbf{I}_3-\bm{\Psi}\cdot \mathbf{Z})\times (\mathbf{I}_3-\bm{\Psi}\cdot \mathbf{Z})^{-1}\,.
\label{Eq_interm_2}
\end{equation}
Finally, inserting Equations~\ref{Eq_interm_1} and \ref{Eq_interm_2} in Equation~\ref{Eq_first_derivative} and using $\mathrm{Tr}(\mathbf{A})=\mathrm{Tr}(\mathbf{A}^\text{T})$ and $\mathrm{Det}(\mathbf{A})=\mathrm{Det}(\mathbf{A}^\text{T})$ yield
\begin{align}
\frac{\partial \mathcal{M}_\Gamma}{\partial \mathbf{Z}} = \left[\mathrm{Det}(\mathbf{I}_3- \mathbf{Z}\cdot \bm{\Psi})\right]^{-\kappa} \exp\!\left[\mathrm{Tr}\!\left(\left[(\mathbf{I}_3- \mathbf{Z}\cdot \bm{\Psi})^{-1}-\mathbf{I}_3\right]\cdot\bm{\Theta}\right)\right] \times \bm{\Psi} \cdot (\mathbf{I}_3- \mathbf{Z}\cdot \bm{\Psi})^{-1} \cdot \left[ \kappa \mathbf{I}_3 + \bm{\Theta}\cdot (\mathbf{I}_3- \mathbf{Z}\cdot \bm{\Psi})^{-1} \right]\,,
\end{align}
or alternatively
\begin{equation}
\frac{\partial\mathcal{M}_\Gamma}{\partial\mathbf{Z}}=\mathcal{M}_\Gamma(\mathbf{Z})\times\bm{\Psi}\cdot(\mathbf{I}_3-\mathbf{Z}\cdot\bm{\Psi})^{-1}\cdot\left[\kappa\mathbf{I}_3+\bm{\Theta}\cdot(\mathbf{I}_3-\mathbf{Z}\cdot\bm{\Psi})^{-1}\right]\, ,
\label{Eq_first_derivative_2}
\end{equation}
using Equation~\ref{Eq_moment_generating_function_gamma}. Therefore, one retrieves the average diffusion tensor of the nc-mv-Gamma distribution (see Equation~\ref{Eq_average_diffusion_tensor_Gamma}):
\begin{equation}
\langle\mathbf{D}\rangle=\left.\frac{\partial\mathcal{M}_\Gamma}{\partial\mathbf{Z}}\right|_{\mathbf{Z}=\mathbf{0}}=\bm{\Psi}\cdot\left[\kappa\mathbf{I}_3+\bm{\Theta}\right] \, .
\label{Eq_average_D_appendix}
\end{equation}



\subsection{Covariance tensor} 

Let us write the first-order matrix derivative of $\mathcal{M}_\Gamma$ Equation~\ref{Eq_first_derivative_2} as
\begin{equation}
\frac{\partial\mathcal{M}_\Gamma}{\partial\mathbf{Z}}=\mathcal{M}_\Gamma(\mathbf{Z})\times\mathbf{F}(\mathbf{Z})\,,
\end{equation}
where
\begin{equation}
\mathbf{F}(\mathbf{Z})=\bm{\Psi}\cdot(\mathbf{I}_3-\mathbf{Z}\cdot\bm{\Psi})^{-1}\cdot\left[\kappa\mathbf{I}_3+\bm{\Theta}\cdot(\mathbf{I}_3-\mathbf{Z}\cdot\bm{\Psi})^{-1}\right]\,,
\end{equation}
with $\mathbf{F}(\mathbf{Z}=\mathbf{0})=\bm{\Psi}\cdot\left[\kappa\mathbf{I}_3+\bm{\Theta}\right]=\langle\mathbf{D}\rangle$ from Equation~\ref{Eq_average_D_appendix}. One can now use the scalar product rule of Equation~\ref{Eq_scalar_product_rule} to obtain
\begin{equation}
\frac{\partial^2\mathcal{M}_\Gamma}{\partial\mathbf{Z}^2}=\frac{\partial[\mathcal{M}_\Gamma(\mathbf{Z})\times\mathbf{F}(\mathbf{Z})]}{\partial\mathbf{Z}}=\frac{\partial\mathcal{M}_\Gamma}{\partial\mathbf{Z}}\otimes\mathbf{F}(\mathbf{Z})+\mathcal{M}_\Gamma(\mathbf{Z})\times\frac{\partial\mathbf{F}}{\partial\mathbf{Z}}\,,
\end{equation}
so that
\begin{equation}
\langle\mathbf{D}^{\otimes 2}\rangle=\left.\frac{\partial^2\mathcal{M}_\Gamma}{\partial\mathbf{Z}^2}\right|_{\mathbf{Z}=\mathbf{0}}=\langle\mathbf{D}\rangle^{\otimes 2}+\left.\frac{\partial\mathbf{F}}{\partial\mathbf{Z}}\right|_{\mathbf{Z}=\mathbf{0}}
\end{equation}
and
\begin{equation}
\mathbb{C}=\langle\mathbf{D}^{\otimes 2}\rangle-\langle\mathbf{D}\rangle^{\otimes\!\,2}=\left.\frac{\partial\mathbf{F}}{\partial\mathbf{Z}}\right|_{\mathbf{Z}=\mathbf{0}}\, .
\label{Eq_CF}
\end{equation}
Using Ref.~\onlinecite{Laue_website} and transposing its result according to Section~\ref{Sec_matrix_calculus}, one has
\begin{align}
\frac{\partial \mathbf{F}}{\partial \mathbf{Z}} & = \kappa \left[\bm{\Psi}\cdot (\mathbf{I}_3-\mathbf{Z}\cdot \bm{\Psi})^{-1} \otimes (\mathbf{I}_3- \bm{\Psi}\cdot \mathbf{Z})^{-1}\cdot \bm{\Psi} \right] \nonumber \\
 & \quad + \bm{\Psi}\cdot (\mathbf{I}_3-\mathbf{Z}\cdot \bm{\Psi})^{-1}\cdot \bm{\Theta}\cdot (\mathbf{I}_3-\mathbf{Z}\cdot \bm{\Psi})^{-1} \otimes (\mathbf{I}_3- \bm{\Psi}\cdot \mathbf{Z})^{-1}\cdot \bm{\Psi} \nonumber \\
 & \quad + \bm{\Psi}\cdot (\mathbf{I}_3-\mathbf{Z}\cdot \bm{\Psi})^{-1} \otimes (\mathbf{I}_3-\bm{\Psi}\cdot\mathbf{Z})^{-1}\cdot \bm{\Theta}\cdot (\mathbf{I}_3-\bm{\Psi}\cdot\mathbf{Z})^{-1}\cdot \bm{\Psi}\,.
\end{align}
In particular,
\begin{equation}
\left.\frac{\partial\mathbf{F}}{\partial\mathbf{Z}}\right|_{\mathbf{Z}=\mathbf{0}}=\kappa\,\bm{\Psi}\otimes\bm{\Psi}+(\bm{\Psi}\cdot\bm{\Theta})\otimes\bm{\Psi}+\bm{\Psi}\otimes(\bm{\Theta}\cdot\bm{\Psi})\,.
\label{Eq_interm_FZ}
\end{equation}
Finally, combining Equations~\ref{Eq_CF} and \ref{Eq_interm_FZ} yields the covariance tensor of the nc-mv-Gamma distribution (see Equation~\ref{Eq_covariance_matrix_Gamma}):
\begin{equation}
\mathbb{C}=\kappa\,\bm{\Psi}\otimes\bm{\Psi}+(\bm{\Psi}\cdot\bm{\Theta})\otimes\bm{\Psi}+\bm{\Psi}\otimes(\bm{\Theta}\cdot\bm{\Psi}) \, .
\end{equation}
This proof is one of the original contributions of the present work.

\end{widetext}

\end{appendix}

\newpage


\end{document}